\documentclass[%
 amsmath,amssymb,
prx,
twocolumn
]{revtex4-2}

\usepackage{graphicx}
\usepackage{dcolumn}
\usepackage{bm}

\usepackage{microtype}

\usepackage{multirow}
\usepackage{booktabs} 
\usepackage{comment}
\usepackage{bbold}
\usepackage{algpseudocode}
\usepackage{amsthm}

\usepackage{siunitx}
\usepackage{xcolor}
\definecolor{myorange}{HTML}{e79534}

\usepackage{relsize}
\usepackage[dvipsnames]{xcolor}
\definecolor{myblue}{RGB}{35, 123, 148}
\definecolor{myred}{RGB}{242, 145, 0}
\usepackage{hyperref}
\hypersetup{
	colorlinks=true,
	linkcolor=myred,
	filecolor=myblue,      
	urlcolor=myblue,citecolor=myblue}
\usepackage{braket}

\usepackage{amsmath}
\usepackage{amssymb}
\usepackage{mathtools}
\usepackage{amsthm}

\usepackage[capitalize]{cleveref}

\begin{document}


\title{Discovering quantum phenomena with Interpretable Machine Learning}

\author{Paulin de Schoulepnikoff}
\email{paulin.de-schoulepnikoff@uibk.ac.at}
\affiliation{University of Innsbruck, Department for Theoretical Physics, Technikerstr. 21a, A-6020 Innsbruck, Austria}
\author{Hendrik Poulsen Nautrup} 
\affiliation{University of Innsbruck, Department for Theoretical Physics, Technikerstr. 21a, A-6020 Innsbruck, Austria}
\author{Hans J. Briegel} 
\affiliation{University of Innsbruck, Department for Theoretical Physics, Technikerstr. 21a, A-6020 Innsbruck, Austria}
\author{Gorka Mu\~noz-Gil} 
\email{gorka.munoz-gil@uibk.ac.at}
\affiliation{University of Innsbruck, Department for Theoretical Physics, Technikerstr. 21a, A-6020 Innsbruck, Austria}

\date{\today}

\begin{abstract}

Interpretable machine learning techniques are becoming essential tools for extracting physical insights from complex quantum data. We build on recent advances in variational autoencoders to demonstrate that such models can learn physically meaningful and interpretable representations from a broad class of unlabeled quantum datasets. From raw measurement data alone, the learned representation reveals rich information about the underlying structure of quantum phase diagrams. We further augment the learning pipeline with symbolic methods, enabling the discovery of compact analytical descriptors that serve as order parameters for the distinct regimes emerging in the learned representations. We demonstrate the framework on experimental Rydberg-atom snapshots, classical shadows of the cluster Ising model, and hybrid discrete-continuous fermionic data, revealing previously unreported phenomena such as a corner-ordering pattern in the Rydberg arrays. These results establish a general framework, which constitutes an important step toward the autonomous discovery of physical descriptors from diverse quantum datasets. All methods are available through \href{https://github.com/qic-ibk/qdisc}{\texttt{qdisc}}, an open-source Python library designed to make these tools accessible to the broader community.

\end{abstract}

\maketitle

\section{Introduction} \label{sec:intro}

The use of machine learning (ML) in physics has grown into a powerful paradigm for uncovering patterns in high-dimensional data, offering new perspectives on longstanding problems in quantum many-body systems~\cite{Neural_Torlai_2018, machine_carrasquilla_2017, mauron2025predicting, Solving_Carleo_2017, bohrdt2019classifying, khosrojerdi2025unsupervised}. Beyond prediction and classification, interpretable ML architectures play a unique role in distilling physically meaningful descriptors of complex quantum data~\cite{Interpretable_Crammer_2023, Interpretable_Suresh_2024, Cybiński_Speak_2024, patel2022unsupervised}. Among these, variational autoencoders (VAEs) have emerged as promising tools for extracting minimal yet informative latent representations directly from raw data~\cite{Learning_Fernandez_2024, moller2025learning}. Recent work has shown that VAEs, when augmented with probabilistic generative decoders, can capture the intrinsic stochasticity of quantum measurements and uncover latent variables that align with physical order parameters~\cite{de2025interpretable}.

Despite this progress, two key challenges remain. First, most existing approaches have been developed and benchmarked primarily on simple spin configurations, and their extension to more general forms of quantum data remains largely unexplored~\cite{walker_deep_2020,Wetzel_Unsupervised_2017, jang_unsupervised_2025, naravane_semisupervised_2023, frohnert2024explainable,de2025interpretable}. In particular, systems with higher local Hilbert space dimension, continuous variables, or data in the form of classical shadows pose nontrivial challenges, as their correlations and statistical structure differ substantially from those of binary spin snapshots. Second, while VAEs can uncover informative latent spaces, attributing clear physical meaning to these representations often relies on post hoc analysis or comparison with known observables, limiting their autonomy in discovering new physics.

In this work, we address both challenges. First, we generalize the probabilistic VAE framework to handle classical shadows as well as hybrid discrete/continuous data of a fermionic system, thereby testing its ability to learn meaningful representations from complex quantum data. Second, to further enhance the interpretability, we integrate symbolic regression techniques into the analysis of latent spaces, enabling the extraction of compact analytic expressions that define the learned clusters. This synthesis of representation learning and symbolic discovery not only uncovers phase structures without prior knowledge but also translates them into compact interpretable physical descriptors.

Our contributions demonstrate that combining probabilistic VAEs with symbolic methods provides a powerful, general, and interpretable framework for representation learning in quantum physics. By successfully applying it to both spin and fermionic systems, we show how such hybrid approaches pave the way toward automated extraction of effective theories from raw quantum data. In particular, we identify a corner-ordering pattern from experimental Rydberg snapshots and, from classical-shadow measurements of the cluster-Ising model, an emergent algebraic behavior that we attribute to finite-size and boundary effects. These results highlight the potential of this hybrid pipeline for the automated discovery of effective descriptions and novel phenomena in complex quantum systems.

\section{Methods} \label{sec:methods}

\begin{figure*}[t]
    \centering
    \includegraphics[width=0.85\textwidth]{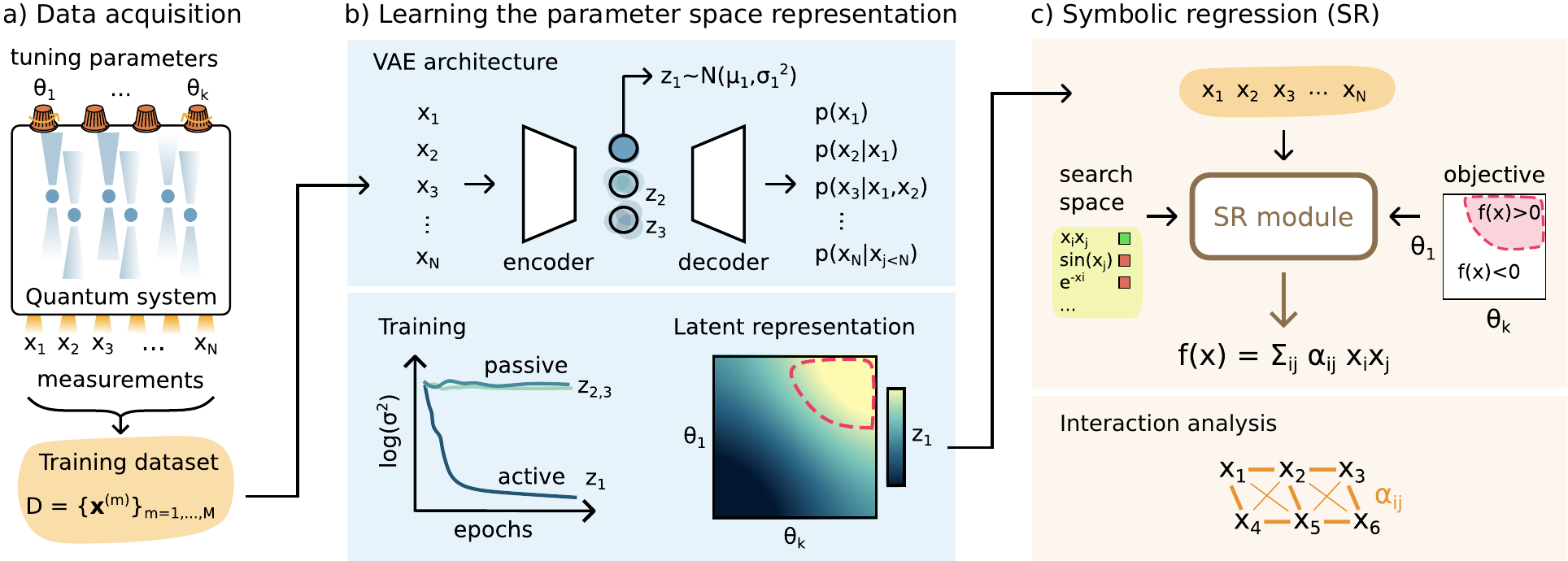}
    \caption{\textbf{Schematic representation of the proposed pipeline.} 
    \textbf{a)} An experimental setup produces data $\mathbf{x}$ of a quantum system for different values of the $k$ experimental parameters $\theta_{1,...,k}$.
    \textbf{b)} A variational autoencoder (VAE) is trained on an ensemble of unlabeled quantum data collected from the previous experiment. It learns to reconstruct the conditional probabilities of its input and encodes the data in few latent variables $\mathbf{z}$. Inspecting the values of the learned latent neurons across the parameter space will provide insight into the system's phase diagram and can reveal unknown features. 
    \textbf{c)} An unknown cluster appearing in the latent representation can be further explained and investigated using symbolic regression methods (SR). The latter seek for a compact analytic formula to discover an order parameter. Its structure can also provide insight into the nature of the cluster and the physics of the system.}
    \label{fig:proposed_pipepline}
\end{figure*}

We now review the main modules of the proposed pipeline, QDisc, schematically represented in~\cref{fig:proposed_pipepline}. QDisc consists of three modules: first, data is acquired from a quantum experiment; second, a variational autoencoder (VAE) is used to generate a parameter space representation; and third, symbolic regression (SR) is used to produce closed-form equations that identify the different regimes observed in learned representations.

In the data acquisition step, we consider a quantum experiment with a certain number of controllable, experimental parameters $\theta_{1,...,k}$ (see~\cref{fig:proposed_pipepline}a), as for instance the strength of a magnetic field, the spacing between atoms in an array or the detuning of some laser. For each set of parameters, we perform a measurement producing a certain string $\mathbf{x}$. The latter can in principle be of any form. In this work, we focus on cases where the subsystems of the experiment, namely the spins, atoms or qubits, can be measured individually in some form, either via discrete projective measurements in a chosen basis or through measurements along random directions, or through continuous local density measurements. In this case, $\mathbf{x} = \{ x_i \}_{i=1:N}$, where $N$ is the number of subsystems. We note that this same pipeline can also be extended to global measurements following~\cite{moller2025learning}.

By acquiring data across a wide range of the experimental parameters $\theta_{1,...,k}$, we generate a dataset $\mathcal{D} = \{ \mathbf{x}^{m} \}_{m=1:M}$, where $M$ is the size of the dataset. By means of this unlabelled dataset, we will train the VAE and subsequently the SR module. We now review in further depth both of these modules. We stress that at no point we consider any prior knowledge on the underlying physics of the experiment explored, and the only information available to us is the measurements $\mathbf{x}^{m}$.

\subsection{Learning parameter spaces via VAE} \label{sec:VAE_background}

To extract interpretable representations from the quantum dataset $\mathcal{D}$, we adopt the VAE framework introduced in Ref.~\cite{de2025interpretable}. The VAE architecture encodes input data into a compact, probabilistic latent space, typically modeled as a multivariate Gaussian distribution. The encoder outputs the mean $\mu_i$ and variance $\sigma_i$ for each latent dimension, from which latent variables $z_i$ are sampled (see Fig.~\ref{fig:proposed_pipepline}b). The resulting latent representation $\mathbf{z} = \{z_i\}_{i=1:Z}$, where $Z$ is the dimensionality of the latent space, is then used to reconstruct the original input data. To deal with the complexity of this task, and account for the rich properties of the input data, we build the encoder based on transformer layers (see \cref{an:NN_archi}).

Given the nature of the quantum data used for training, one must correctly formulate the reconstruction problem. While previous VAE applications in the quantum domain relied on deterministic decoders~\cite{walker_deep_2020, Wetzel_Unsupervised_2017, jang_unsupervised_2025, naravane_semisupervised_2023, frohnert2024explainable}, this formulation inherently clashes with the stochastic nature of measurement-derived quantum data. Instead, one must aim to approximate the underlying distribution of the input data~\cite{Learning_Fernandez_2024}. Therefore, the challenge of classically representing quantum distributions necessitates a decoder architecture capable of closely capturing the properties of quantum states. To this end, we design our decoder based on recent advances in Neural Quantum States (NQS) \cite{Solving_Carleo_2017, From_Lange_2024}, ensuring sufficient expressiveness to faithfully reconstruct the input quantum states.

In practice, we implement the decoder as an autoregressive neural network. Its design follows from the conditional factorization of the joint probability distribution,
\begin{equation}
\label{eq:cp}
p(\mathbf{x}) = \prod_{i=1}^{N} p(x_i \mid x_{i-1}, \ldots, x_1),
\end{equation}
which expresses the full distribution as a product of conditional probabilities. Accordingly, the network predicts the conditional probability of $x_i$ given all preceding elements. Importantly, an ordering for these conditional probabilities must be chosen, in particular for systems of dimension greater than one. The ordering chosen for each system explored below is detailed in~\cref{an:QSyst_details}.

This probabilistic formulation ensures that the model learns the underlying distribution of the input ensemble, allowing it to capture nontrivial many-body correlations beyond mean-field descriptors. Moreover, as we will see in \cref{sec:results_Rybderg}, the direct access to these probabilities represents yet another interpretable signal that we can use to further understand the properties of the data at hand.

The training is performed by minimizing the loss

\begin{equation} \label{eq:loss_VAE}
    \mathcal{L} = -\frac{1}{M}\sum_{m=1}^{M}\mathbb{E}_q[\log p(\mathbf{x}^{m}|\mathbf{z})] - \beta \mathrm{KL}[q(\mathbf{z}|\mathbf{x}^{m})||p(\mathbf{z})].
\end{equation}
The first term encourages faithful probabilistic reconstruction. The second term is a regularization term, which minimizes the Kullback-Leibler (KL) divergence between the latent space distribution $q(z_i|\mathbf{x}^{(m)})\sim \mathcal{N}(\mu_i,\sigma_i^2)$ and the prior $p(z)$, typically defined as standard normal distribution $p(z)\sim \mathcal{N}(0,1)$. 
This term acts as a regularizer, enforcing the latent representation to align with the uninformed prior distribution. The trade-off between these two terms creates a competition, such that only the minimum number of latent neurons required for proper input reconstruction diverge from the prior (referred to as active neurons), while the others collapse to the prior distribution (referred to as passive neurons) (see \cref{fig:proposed_pipepline}b). The hyperparameter $\beta$ is introduced to ensure that the two terms are properly balanced~\cite{Understanding_Burgess_2018}. More details on the loss are presented in~\cref{an:loss}.

This procedure enables the latent space to encode a meaningful representation of the target phase diagram, as demonstrated in Ref.~\cite{de2025interpretable}. Specifically, plotting the values of the \emph{active} latent neurons against the experimental parameters $\theta_{1,\ldots,k}$ reveals distinct regimes in the data (see \cref{fig:proposed_pipepline}b). While such visualizations provide valuable insights into the presence of different phases, our objective is to further enhance the interpretability by deriving closed-form equations that serve as order parameters, thereby distinguishing the identified regimes.

\subsection{Order parameter discovery via symbolic regression}
\label{sec:SR}
Symbolic regression (SR) provides closed-form equations for a specified problem, typically through a genetic-algorithm-based exploration of candidate mathematical terms~\cite{Interpretable_Crammer_2023}. Its ability to distill complex interactions into interpretable equations makes it a powerful tool for uncovering unknown physical phenomena~\cite{hafner2023machine, abdussalam2025symbolic}. Here, we leverage the insights obtained from training the VAE, combined with SR, to identify closed-form order parameters that distinguish the phases of interest encoded in the latent space. Order parameters are essential in quantum systems as they quantitatively identify and distinguish distinct phases by capturing symmetry breaking and emergent properties, while also enabling the characterization of critical behavior near phase transitions. Their measurement and analysis provide both theoretical insight and experimental validation for understanding the fundamental organization of quantum matter.

To achieve this, we first isolate the region of interest in the parameter space using the VAE representation (see circled area in Fig.~\ref{fig:proposed_pipepline}b). We then frame the task as a classification problem, where the goal of symbolic regression (SR) is to generate a function $f(\textbf{x})$, where $\textbf{x}$ are samples of the dataset $\mathcal{D}$, that is positive within the target region and negative outside (see Fig.~\ref{fig:proposed_pipepline}c). For the search space, we consider a broad set of mathematical terms. Starting from a set of elementary operations (e.g., $+$, $-$, $\cdot$, $\exp(x_i)$, $\sin(x_i)$, ...), the underlying genetic algorithm can generate a wide variety of symbolic functions, thereby exploring a large space of mathematical expressions. This space being very large, the search can benefit from prior knowledge on the dataset at hand. In particular, if the relevant expression is expected to involve interactions of at most $k$ bodies, the search space can be restricted to monomials of degree $k$ in the $x_i$, in which case the symbolic search reduces to a polynomial regression over $\sim N^k$ terms. As we will see, these terms can be further compressed using a symbolic search.

Although straightforward, this objective aligns with the fundamental definition of an order parameter. While alternative objectives, such as a recent proposal leveraging the phase boundary's gradient~\cite{wetzel2025closed}, are also valid, the classification objective yielded both robust quantitative performance and highly interpretable equations (benchmark results are presented in \cref{an:SM_on_J1J2}).

\begin{figure*}[t]
    \centering
    \includegraphics[width=1.\linewidth]{figures/VAE4Spins_Rydberg_new.pdf}
    \caption{\textbf{Experimental Rydberg atoms array.}
    \textbf{a)} Phase diagram of the Rydberg-atom system, based on~\cite{Machine_Miles_2023}, together with exemplary ordered configurations on a reduced $5\times 5$ lattice. White, purple, and orange circles indicate atoms in the $\ket{g}$, $\ket{r}$, and $\ket{+_x}$ states, respectively.  
    \textbf{b)} Atoms are measured as being in the ground (g) or Rydberg (r) state, leading to a binary data stream.
    \textbf{c)} Training of the VAE seen through the log of the variances $\log(\sigma_i^2)$ of the latent variables $z_i$ output by the encoder. 
    \textbf{d)} Latent representation learned by the VAE. For the two active latent neurons, we show the mean values $\mu_i$ produced by the encoder for input snapshots sampled across the parameter space.  
    \textbf{e)} Conditional probabilities generated by the decoder of the VAE at several locations in the latent space, indicated in panel~d). 
    \textbf{f)} Top: location of the snapshots used to construct the classification dataset for SR. Middle: prediction of the discovered function $f$. Bottom: prediction of the order parameter $\tilde{f}$ constructed by subtracting correlations on the edges and the bulk. The red dashed curve highlights the regions of interest.
    }
    \label{fig:Rydberg}
\end{figure*}

\subsection{QDisc: a python library for Quantum phenomena Discovery}

To facilitate the practical use of the proposed pipeline, we developed \verb|qdisc|, a dedicated Python library for \textbf{Q}uantum phenomena \textbf{Disc}overy ~\cite{github_of_the_project}. The library’s structure closely follows the stages of the pipeline presented here and includes a variety of examples and tutorials that reproduce key results from this work.

The library comprises three core modules. First, \verb|qdisc.Dataset| enables users to handle and preprocess their data efficiently. Next, \verb|qdisc.vae| provides tools to construct, customize, and train variational autoencoders (VAEs), including predefined architectures and high-level training modules. Finally, \verb|qdisc.sr| allows users to train symbolic regression (SR) models based on insights gained from the VAEs. Designed as an accessible entry point, \verb|qdisc| empowers users, even those with minimal machine learning experience, to leverage the tools developed in this study.

\section{Results}\label{sec:results}

In the following section, we apply QDisc to a range of paradigmatic quantum experiments, demonstrating the versatility and power of our pipeline. First, we analyze experimental data from a Rydberg atom array~\cite{ebadi_quantum_2021}, revealing QDisc’s ability to uncover previously unknown regimes. Next, we assess its performance on inputs from a random measurement scenario~\cite{elben2023randomized} on the Cluster Ising model, revealing an unexpected regime within its symmetry-protected phase. Finally, we extend our method to hybrid data arising from a dual-species interacting fermionic model~\cite{falicov1969simple}, where one species is probed via discrete occupation measurements and the other via continuous particle density measurements.

\subsection{QDisc identifies a corner ordered-regime in experimental Rydberg arrays} \label{sec:results_Rybderg}

Rydberg atom arrays have emerged as one of the most promising platforms for quantum simulation and computation~\cite{browaeys2020many,anand2024dual,celi2020emerging,cesa2023universal}. In this section, we analyze data from the experiment reported in Ref.~\cite{ebadi_quantum_2021}, which investigates the phases arising in a square lattice of Rydberg atoms. The system is characterized by two tunable experimental parameters: the lattice spacing $a$, which controls the effective range of the Rydberg blockade $R_b$ relative to the lattice geometry, and the detuning $\Delta$, which sets the energy offset between the ground state and Rydberg state, thereby acting as an effective longitudinal field. The resulting phases in the explored parameter space are schematically illustrated in \cref{fig:Rydberg}a. More details on the system are presented in~\cref{an:Rydberg_details}.

Binary snapshots are collected across the entire parameter space, representing atoms in either the ground state or the Rydberg state (see \cref{fig:Rydberg}b). This dataset is then used to train the VAE. During training, we observe that two neurons remain active, while all others collapse to the prior (see \cref{fig:Rydberg}c), as evidenced by their variances $\sigma^2$ decreasing, while the inactive neurons converge to $\sigma^2 \rightarrow 1$. This indicates that two latent dimensions suffice to describe the dataset. Notably, one latent neuron, $z_2$, exhibits a significantly larger variance ($\sigma_2^2$) than the other. This behavior suggests that the neuron $z_1$ encodes more fine-grained information, allowing for less noise injection (lower variance) to ensure accurate reconstruction by the decoder. We note that this interpretation assumes that the active neurons encode information the decoder needs, i.e. the latter cannot properly reconstruct without accessing them \cite{chen2016variational}. To ensure this, one can for instance analyse how the reconstructed samples vary when a specific active latent neuron is changed \cite{de2025interpretable}.

Inspecting the mean values $\mu_i$ of the neurons (see \cref{fig:Rydberg}d), we find that $z_1$ effectively distinguishes most of the experiment's phases, serving as the primary information source for the decoder, which is reflected by its smaller $\sigma^2_1$. In contrast, $\mu_2$ captures more subtle features, particularly highlighting the striated phase ($\mu_2 \sim -1$). Additionally, this neuron reveals an unexpected cluster ($\mu_2 > 1$) within the edge-ordered phase, not identified by previous ML approaches~\cite{Machine_Miles_2023}, underscoring the VAE's ability to uncover unknown regimes directly from raw quantum data.

In order to uncover the physical nature of the identified regime, we first inspect the conditional probabilities (\cref{eq:cp}) when sampling latent vectors at different parts of the parameter space (\cref{fig:Rydberg}e). Starting from the empty phase (at location 1) with all probabilities close to zero, we see that the ordering appears first at the corners (location 2) and at the edges (location 3) to then extend in the bulk (location 4).

We now apply symbolic regression (SR), leveraging the insights gained from the previous analysis to guide its optimization. In particular, by analyzing the conditional probabilities, we identify a relevant ordering predominantly occurring at the corners of the lattice. This differs from the expected edge-ordering. Hence, we aim to find an order parameter that can differentiate between the newly found regime and the edge-ordered phase. To do so, based on the previous analysis, we restrict the search space to two-body terms of the form $\alpha_{ij} x_i x_j$, where $i,j$ label the four corner sites (labels $i,j=1-4$ denote the corner, adjacents and opposite atoms, respectively) and $\alpha_{ij}$ are the corresponding coefficients. To define the classification objective, we set as positive the points within the identified cluster (upper panel in~\cref{fig:Rydberg}f). As we are here interested in differentiating this regime from the edge-ordered phase, we consider as negative only points identified as the latter. Moreover, because the boundary is not perfectly defined in the latent space (\cref{fig:Rydberg}d), we consider for the classification problem points sufficiently far from the boundary.

SR then finds an order parameter, $f(\mathbf{x})=x_1x_4+2x_2x_3$, which correctly solves the classification problem (central panel in \cref{fig:Rydberg}f). A robustness analysis is presented in \cref{an:robustness_symb_expr}. Nonetheless, the learned function is also positive within the star and checkerboard phase. Indeed, the SR was not optimized with configurations from these phases, and hence the shown outcome is the result of generalization. Nonetheless, discarding these phases is simple: as these are bulk ordered phases, we subtract from $f$ the nearest-neighbor correlator $\langle \sigma_i^z\sigma_j^z\rangle$ for atoms in the bulk and edges, finding an order parameter $\tilde{f}$ that can uniquely distinguish the newly identified regime (bottom panel in \cref{fig:Rydberg}f). From a physical perspective, it is known that ordering in this experimental platform is largely boundary-driven. The ordered structures typically emerge from the edges and progressively extend toward the bulk of the system, enabling the access of phases that would otherwise be hidden behind a first-order transition line~\cite{kalinowski2022bulk}. Given the major role of boundaries, it is therefore natural to find a regime in which the ordering is developed only at the corners, as these are the regions with the greatest boundary effect.

\begin{figure*}
    \centering
    \includegraphics[width=1.\linewidth]{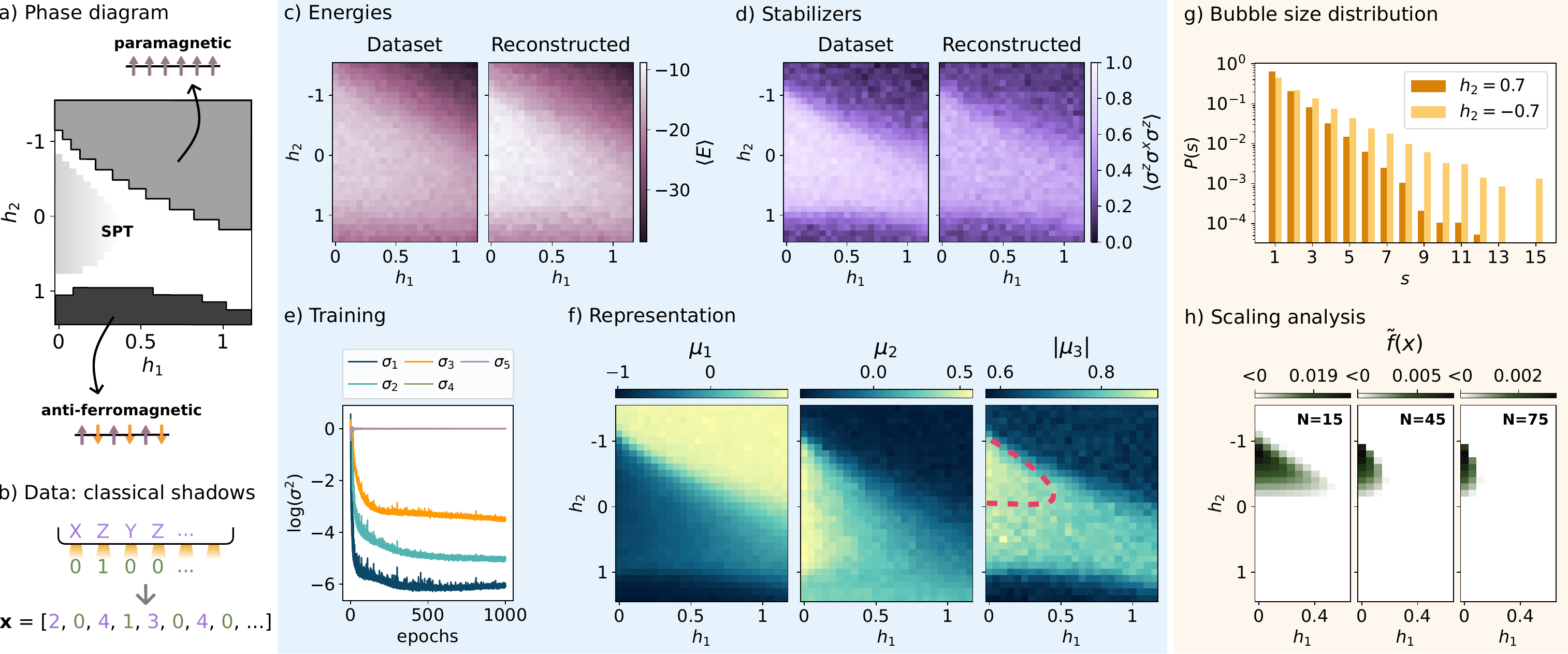}
    \caption{\textbf{Shadows of the cluster Ising model.} 
     \textbf{a)} Phase diagram of the cluster Ising model, based on the values of the nearest-neighbor correlator and the string order parameter. To illustrate the gradient of the latter caused by finite-size effects, the area with high value is highlighted in light grey.
     \textbf{b)} Spins are measured in a random basis, leading to a data stream containing the basis and the measurement outcome for each spin.
    \textbf{c)} Expectation values of the energy computed from the shadows of the dataset and reconstructed by the VAE.
    \textbf{d)} Expectation values of the three-body stabilizer $\sigma_{i-1}^z \sigma_i^x \sigma_{i+1}^z$ computed from the shadows of the dataset and reconstructed by the VAE.
    \textbf{e)} Training of the VAE seen through the log of the variances $\log(\sigma_i^2)$ of the latent variables $z_i$ output by the encoder. 
    \textbf{f)} Latent representation learned by the VAE.
    \textbf{g)} Distribution $P(s)$ of $X$-bubbles as a function of their size $s$ for $h_1=0.05$ and $h_2=\pm 0.7$.
    \textbf{h)} Prediction of the normalized symbolic expression $\tilde{f}$ across the parameter space for chains of $N=15,45$ and $75$ spins.}
    \label{fig:clusterIsing}
\end{figure*}

\subsection{QDisc uncovers power-law scaling in a random measurements setting} \label{sec:cluster_ising}

In many quantum systems, a single measurement basis is insufficient to fully characterize the underlying state. Randomized measurements have emerged as a resource-efficient alternative to costly full quantum tomography~\cite{elben2023randomized}, and classical shadows have been shown to enable provably efficient ML-based phase classification~\cite{huang2020predicting, huang2022provably}. Here, we leverage the classical shadow framework to explore the parameter space of the Cluster Ising spin model, a paradigmatic example exhibiting symmetry-protected topological phases~\cite{smacchia2011statistical, verresen2017one}.
To explore this model, we perform random $X$, $Y$, and $Z$ measurements on each spin. The resulting data, known as the classical shadow, is encoded as a $2N$-dimensional vector that records both the measurement basis and the outcome for each spin (see \cref{fig:clusterIsing}b). We generate a dataset by collecting shadows across the model's parameter space, which is parameterized by two variables: a transverse-like field $h_1$ and an Ising coupling $h_2$ (see \cref{an:clusterIsing_details} for more details).

We then train the VAE on this dataset. Before analyzing the learned representation, we demonstrate the VAE's ability to accurately capture the underlying physics of the model, even in this randomized measurement setting. To validate this, we compare two observables, the configuration’s energy and three-body stabilizer $\sigma_{i-1}^z \sigma_i^x \sigma_{i+1}^z$, computed both on the training dataset and on the outputs generated by the VAE given some random measurement basis (see \cref{fig:clusterIsing}c and d). We then compute expectation values for both observables using the method of moments (see \cref{an:classical_shadows}). These results show that the VAE successfully generates shadows with the expected physical properties, even for complex observables such as the three-body stabilizer, which is only slightly underestimated in the SPT phase. 

We now analyze the learned latent representation. 
While the system's Hamiltonian has only two tunable parameters, three latent neurons remain active (see~\cref{fig:clusterIsing}e). Although increasing the regularizer $\beta$ in \cref{eq:loss_VAE} reduced the number to two, this significantly degraded the reconstruction accuracy, while leaving the additional cluster resolved (see \cref{an:cluster_Ising_add_res} for more details). We therefore decided to retain three active neurons. We attribute this to the complexity of the randomized dataset, which challenges the encoder's ability to identify a minimal representation. Nevertheless, all three neurons encode meaningful information: the first captures the boundaries between the three phases, the second reflects a gradient across the SPT phase, consistent with the expected behavior of the string order parameter in finite-size systems~\cite{venuti2005analytic}. Because the $\mathbb{Z}_2\times\mathbb{Z}_2$ symmetry causes the mean of $\mu_3$ to vanish on average, we plot its average absolute value $|\mu_3|$, revealing an unexpected region at the edge of the SPT phase.

To investigate this region, we employ the SR classification scheme with a search space that includes all two-body interaction terms $\alpha_{ij}x_i x_j$.
To avoid the complexity of processing the random measurement scenario, we restrict this analysis to configurations obtained from a fixed measurement pattern. In particular, we find that configurations arising from $X$ measurements allow the symbolic regression procedure to identify an expression that successfully solves the classification task (see \cref{an:cluster_Ising_add_res}).
The optimized weights show an interplay between short- and long-range interactions. To elucidate this, we apply a genetic symbolic algorithm~\cite{Interpretable_Crammer_2023} to derive a closed-form expression $g(i,j) = \alpha_{ij}$, leading to the order parameter
\begin{align} \label{eq:final_f_cluster_Ising}
    f(\mathbf{x}) = \sum_{i,j>i} \underbrace{\Big(-0.20 + \frac{0.07}{r_{ij} - 0.95} \Big)}_{\displaystyle g(i,j)} \; x_i \cdot x_j,
\end{align}
where $r_{ij} = j-i$ is the distance between spins. A robustness analysis is presented in \cref{an:robustness_symb_expr}. This symbolic expression provides interesting insights into the underlying physics of this extra cluster. Indeed, the coefficients $g(i, j)$ are positive only for nearest neighbors (NN) and negative otherwise, with long-range dependencies on the distance $r$ algebraically. 

Furthermore, competing short-range ferromagnetic and long-range antiferromagnetic interactions are known to give rise to periodic stripe patterns, as well as island or bubble-like structures in certain regimes~\cite{giuliani2016periodic}. We define an $X$-bubble of size $s$ as a connected group of $s$ consecutive spins $x_i, \dots, x_{i+s-1}$ ferromagnetically aligned along the $X$ direction. This motivates the analysis of the distribution $P(s)$ of $X$-bubble sizes. As shown in~\cref{fig:clusterIsing}g, within the additional cluster $(h_1,h_2)=(0.05,-0.7)$, the distribution $P(s)$ exhibits a slower decay with increasing $s$ compared to the mirrored location $(h_1,h_2)=(0.05,0.7)$. Additional analysis, considering points throughout the entire SPT phase, are presented in \cref{an:cluster_Ising_add_res}.

To validate the order parameter, we perform a scaling analysis of the normalized $\tilde{f}(\mathbf{x}) = f(\mathbf{x})/[0.5N(N-1)]$ for larger chains of length $N=45$ and $N=75$ using density-matrix renormalization group (DMRG) calculations~\cite{white1992density}. The results (with a truncation error below $4.5\times10^{-7}$) show that the region where $\tilde{f}(x) > 0$ progressively shrinks along the $h_1$ direction, while no comparable shrinkage is observed along $h_2$ (see \cref{fig:clusterIsing}h). In addition, the maximal value of $\tilde{f}(\mathbf{x})$, located around $h_2 \simeq -0.7$, decreases with increasing system size, indicating that this feature becomes less pronounced in the thermodynamic limit and may instead originate from finite-size effects associated with phase boundaries. Further investigation of this behavior is beyond the scope of this work and will be addressed in future studies.

\subsection{QDisc uncovers fermion-fermion interactions from hybrid discrete-continuous quantum data} \label{sec:fkm}

\begin{figure*}[t]
    \centering
    \includegraphics[width=0.85\linewidth]{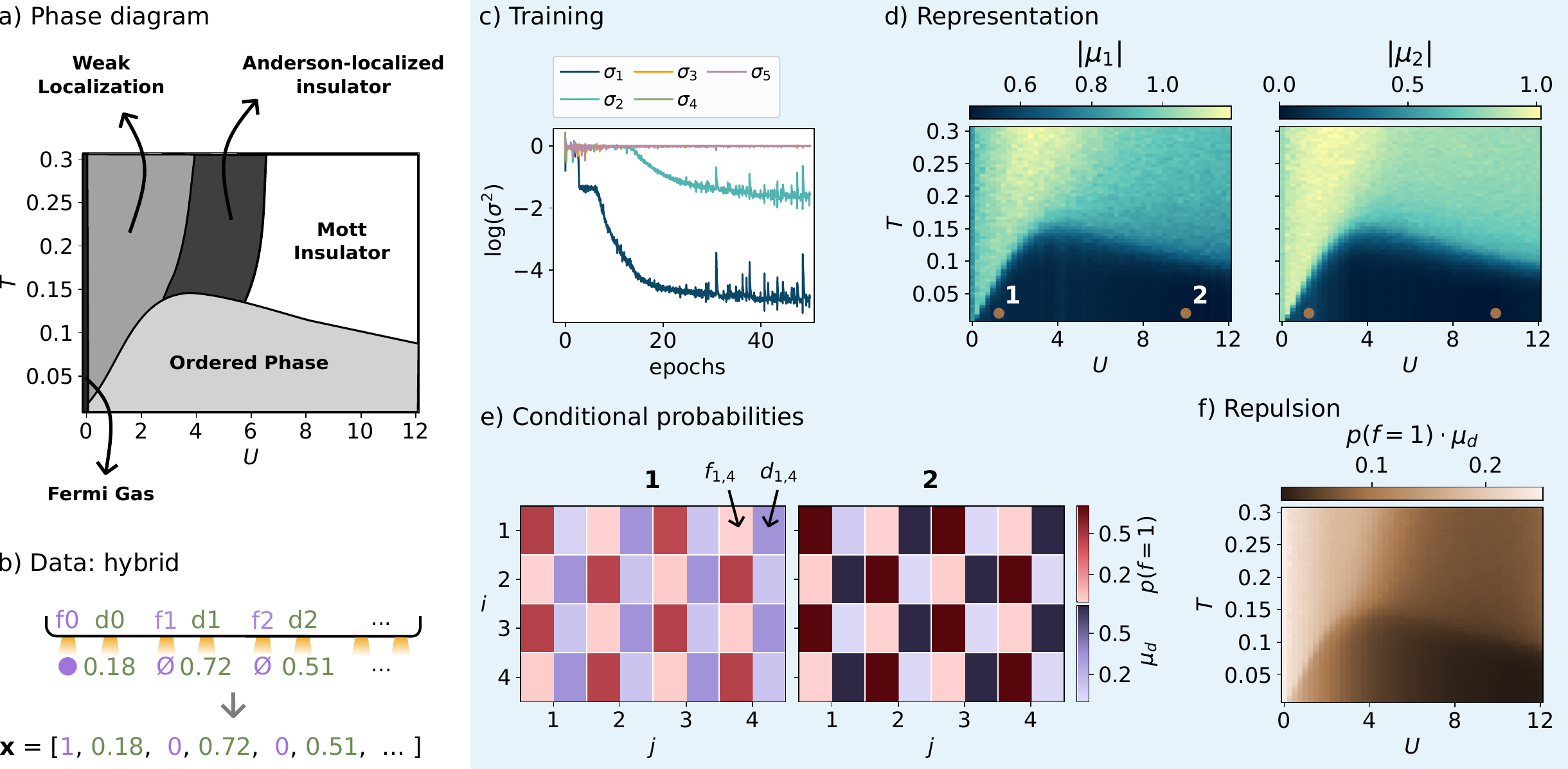}
    \caption{\textbf{Hybrid data of the Falicov-Kimball model.} 
     \textbf{a)} Phase diagram of the FKM, based on~\cite{antipov2016interaction, frk2025unsupervised}.
     \textbf{b)} The two fermion species are measured, leading to a binary outcome (occupied or empty) for $f$ fermions and a continuous local density for $d$ fermions.
    \textbf{c)} Training of the VAE seen through the log of the variances $\log(\sigma_i^2)$ of the latent variables $z_i$ output by the encoder. 
    \textbf{d)} Latent representation learned by the VAE seen through the average of the absolute mean $|\mu_{1,2}|$ across the parameter space.
    \textbf{e)} Conditional probabilities generated by the decoder of the VAE at two different locations in the ordered phase.
    \textbf{f)} The repulsion of $f$--$d$ particles across the parameter space is seen through the product of the conditional probabilities of each species.
    }
    \label{fig:FKM}
\end{figure*}

So far, we have focused on measurements with discrete outcomes. However, many quantum experiments rely on continuous measurements. While purely continuous signals, such as interference patterns or time-of-flight images in quantum simulators~\cite{moller2025learning, eberz2023detecting}, can also be analyzed, we concentrate here on onsite continuous measurements. Specifically, we consider the spinless Falicov--Kimball model (FKM) at half filling on a square lattice~\cite{falicov1969simple}, where localized $f$ particles interact on-site with itinerant $d$ fermions via a repulsive interaction $U$. Crucially, the $f$ fermions are measured using discrete occupation measurements, while the $d$ fermions are characterized by their local densities, a continuous observable. By exploring the full parameter space (sketched in \cref{fig:FKM}a), spanned by the interaction strength $U$ and temperature $T$, we observe several distinct regimes: a checkerboard ordered phase, a Fermi gas, a weak localization of the $f$ particles, an Anderson-like insulator regime, and a strongly correlated Mott insulating phase~\cite{antipov2016interaction, frk2025unsupervised}. Further details on the physical model can be found in~\cref{an:FKM_details}.

From a practical perspective, the VAE must be adapted to handle this hybrid data. While the same transformer encoder used in previous sections can directly process such inputs, we design the decoder to predict both the occupations of the $f$ particles and the local densities of the $d$ particles. Specifically, the decoder produces a sequence in which the odd components correspond to the conditional probabilities of $f$ particle occupation, while the even components encode the parameters describing the $d$ particle densities (see \cref{fig:FKM}b). The latter are modeled through a Gaussian conditional distribution,
\begin{equation}
p(x_{2i} \mid x_{j<2i}) = \mathcal{N}\big(\mu_d, \sigma_d^2\big),
\end{equation}
where the mean $\mu_d$ and standard deviation $\sigma_d$ are predicted by the decoder given the preceding elements $x_{j<2i}$ of the sequence. 

Training the VAE with hybrid configurations across the parameter space leads to two active latent neurons (see~\cref{fig:FKM}c). Due to the symmetry imposed by half filling, the latent variable means vanish throughout the parameter space. Therefore, we plot the average of the absolute values of the latent means, shown in~\cref{fig:FKM}d. The ordered phase clearly emerges in the latent representation. In addition, several subregions become distinguishable, including the Fermi-gas, the Mott-insulator, and the weak-localization regimes. The value of $|\mu_2|$ also gradually varies on the left side of the ordered phase.

To investigate the physical origin of this variation, we leverage the conditional probabilities produced by the VAE decoder. These are shown in~\cref{fig:FKM}e for $T=0.01$ and interaction strengths $U=0.1$ and $U=10$ (see~\cref{fig:FKM}d for the corresponding locations in parameter space). In this plot, the red element at position $ij$ corresponds to the predicted occupation probability of an $f$ particle, while blue encodes the predicted mean density $\mu_d$ of a $d$ particle at the same site. To account for the $\mathbb{Z}_2$ symmetry, we only used input configurations in which the site $(1,1)$ was occupied by an $f$ fermion. 

For both particle species, the checkerboard pattern characteristic of the ordered phase is visible. In addition, an effective repulsion between $f$ and $d$ particles can be observed: when a lattice site exhibits a high probability of hosting an $f$ particle, the corresponding $d$ particle density is suppressed, and vice versa. This anti-correlation becomes more pronounced as the interaction strength $U$ increases (location 1 versus location 2). Computing the product of probabilities of each species, $\mathrm{p(f_{ij}=1)\cdot p(\mu_{dij})}$, provides a visualization of the interspecies repulsion across the entire parameter space (see~\cref{fig:FKM}f). 

These observations indicate that the variation of $\mu_2$ does not correspond to a distinct thermodynamic phase. Instead, it subdivides the ordered phase according to the strength of the effective $f$-$d$ repulsion, i.e., how strongly an $f$ occupation suppresses the $d$ densities on the same lattice site. Again, this analysis highlights the interpretable nature of the VAE's latent space and the output probabilities, showcasing their strength at capturing subtle physical phenomena. The use of symbolic methods also supports this interpretation (see~\cref{an:SR_FKM}).

\section{Discussion}\label{sec:discussion}

In this work, we have introduced a general pipeline for the discovery and characterization of phases directly from raw quantum data. Building on the variational auto-encoder framework, the method learns a compact and physically meaningful latent representation that aligns with the underlying structure of the phase diagram. Importantly, the approach remains agnostic to prior physical assumptions, enabling the identification of previously unknown regimes. To complement this representation learning step, we incorporated a symbolic regression module designed to extract compact and interpretable analytical descriptors that characterize previously unknown structures. Although genetic-algorithm-based symbolic regression, in principle, allows for the exploration of very large functional spaces, our results suggest that incorporating physically motivated constraints in the search space can be beneficial, both in terms of interpretability and stability of the discovered expressions. In the current framework, such constraints are provided by the user, guided by the information extracted by the VAE but also their own prior knowledge.

We demonstrated the versatility of the pipeline across different types of quantum data, including experimental snapshots of Rydberg atom arrays, classical shadows of the cluster Ising model, and hybrid discrete--continuous data from a fermionic system. In each case, the method successfully captured the main physical regimes and revealed additional structures. Notably, it identified a corner-ordering regime in the Rydberg platform and signatures of algebraic behavior in the cluster Ising model, highlighting its ability to uncover and characterize unknown features. While some of these features may originate from finite-size effects, their detection already demonstrates the pipeline's capacity to uncover subtle and unexpected phenomena. 

While the framework is well positioned to handle the larger datasets produced by next-generation quantum simulators, its reach is fundamentally limited by the information content of the measurements: no learning procedure can identify descriptors that are not present in the data. For instance, a snapshot taken in a single basis cannot encode off-diagonal properties of the state. This would require multi-basis protocols, such as the classical shadows employed in \cref{sec:cluster_ising}, or handling more than one single-basis snapshot at a time \cite{sotnikov2022certification, xiao2024extracting}. More broadly, identifying the right measurements and data representations for exotic phases, such as superconductivity \cite{yang1962concept, hartke2023direct} or topological order \cite{semeghini2021probing}, remains an active and important challenge for data-driven approaches. 

In typical experimental workflows, parameter regimes are often explored based on prior theoretical expectations. By contrast, the present framework enables a more hypothesis-free exploration. That means, by training the full pipeline with data acquired in regions of parameter space whose phase structure is unknown in advance, it can facilitate the discovery of unexpected phenomena. While the generalization capabilities of VAEs may enable the characterization of parameter spaces from sparsely sampled data~\cite{de2025interpretable}, identifying phases of matter absent from the training dataset remains a key open problem for this and other generative approaches to quantum data~\cite{rende2025foundation,RydbergGPT_Fitzek_2024}. Similarly, an order parameter $f(\mathbf{x})$ learned by the SR module only separates the two regions defined in its classification dataset and should not be expected to distinguish other phases.

Looking ahead, the pipeline could be integrated more tightly with experimental platforms. For instance, embedded within a reinforcement learning loop, an AI model could actively guide the exploration of the parameter space toward regions exhibiting novel or highly structured behavior~\cite{melnikov2018active}. Having such an embedded strategy could also improve the representation learned by the VAE~\cite{munoz2026disentanglement}. Such a closed-loop approach would provide a promising route toward autonomous discovery in quantum many-body experiments.

\section*{Acknowledgments}

We sincerely thank Hannes Pichler, Sepehr Ebadi, Tout Wang, and Mikhail Lukin for providing the experimental data on the Rydberg atoms.

This research was funded in part by the Austrian Science Fund (FWF) [SFB BeyondC F7102, DOI: 10.55776/F71; WIT9503323, DOI: 10.55776/WIT9503323]. For open access purposes, the authors have applied a CC BY public copyright license to any author accepted manuscript version arising from this submission. This work was also supported by the European Union (ERC Advanced Grant, QuantAI, No. 101055129). The views and opinions expressed in this article are however those of the author(s) only and do not necessarily reflect those of the European Union or the European Research Council - neither the European Union nor the granting authority can be held responsible for them.

\section*{Data availability}

The data and code used to produce the results of this paper are available through the Python \verb|qdisc| library~\cite{github_of_the_project}. On the other hand, the dataset for the cluster Ising model was created by means of NetKet~\cite{Netket_Filippo_2022, Codebase_Filippo_2022}, which is based on Jax~\cite{JAX_James_2018} and MPI4Jax \cite{mpi4jax_Dion_2021}. The neural networks have been implemented using Flax~\cite{Flax_Heek_2024}.  The genetic searches were performed using PySR~\cite{Interpretable_Crammer_2023}.

\section*{Competing interests}
The authors declare no competing interests.

\section*{Author contributions}
Conceptualization: P.d.S., H.P.N., H.J.B and G.M.-G. Methodology: P.d.S., H.P.N. and G.M.-G.  Software: P.d.S. Formal analysis: P.d.S., H.P.N., H.J.B and G.M.-G. Writing: P.d.S., H.P.N., H.J.B and G.M.-G.

\bibliography{bibiography}

\newpage

\onecolumngrid

\appendix
\crefalias{section}{appendix}
\crefalias{subsection}{subappendix}
\crefalias{subsubsection}{subsubappendix}

\renewcommand{\thefigure}{A\arabic{figure}}
\renewcommand{\thetable}{A\arabic{table}}
\setcounter{figure}{0}
\setcounter{table}{0}
\renewcommand{\theHfigure}{A\arabic{figure}}
\renewcommand{\theHtable}{A\arabic{table}}

\newpage

\section{Neural network details} \label{an:NN_archi}

In this appendix, we provide further details on the architecture of the neural network used for the encoder and decoder of the VAE through the different scenarios presented.

For all numerical simulations presented in the main text, we used a transformer architecture for the encoder and decoder~\cite{Vawani_Attention_2017}. Details are presented in~\cref{tab:archi_TransfEncoder} and~\cref{tab:archi_TransfDecoder} for the encoder and decoder, respectively. For the latter, context based attention is used with the context being the latent variables $\mathbf{z}$. For the decoder to model the conditional probabilities, an ordering must be defined. This depends on the quantum system and is detailed in~\cref{an:QSyst_details}.

For the decoder used for the fermionic system (\cref{sec:fkm}), instead of a single projection head (last two rows of~\cref{tab:archi_TransfDecoder}: \textit{Dense} + \textit{softmax}), two independent projection heads are used for the conditional probabilities of each particle species. These consist of two dense layers with a ReLU activation function between them. Instead of the softmax activation used for the rest of the cases, we use a sigmoid activation to predict the mean densities of $d$ particles. 

The expressivity of the transformer enabled us to use the same architecture for all simulations of the main text, keeping our approach general and unbiased from prior knowledge. We note, however, that prior knowledge such as symmetries, correlation length, or topology could enhance performance and reduce the number of parameters \cite{malyshev2023autoregressive, viteritti2026approaching, yang2020scalable}.

We note that tuning the expressivity of the decoder, for instance by restricting its receptive field as in WaveNet-style architectures, could influence the learned representation \cite{chen2016variational}. Investigating this effect systematically is an interesting direction left for future work.

\begin{table}[h]
    \centering
    \begin{tabular}{l|c}
    \hline
    \hline
       Layer type  & Output size \\
       \hline
       Input &  $B\times I $ \\
       Embedding  & $B\times I \times d_{\mathrm{model}}$ \\
       Positional Encoding ($\sin$) & $B\times I \times d_{\mathrm{model}}$ \\
       Transformer layer (3x):  &  \\
       \hspace{0.5cm} Normalization & $B\times I \times d_{\mathrm{model}}$ \\
       \hspace{0.5cm} Self-Attention ($N_{head}=2/\textcolor{myred}{4}$) & $B\times I \times d_{\mathrm{model}}$ \\
        \hspace{0.5cm} Residual Connection & $B\times I \times d_{\mathrm{model}}$ \\
        \hspace{0.5cm} Dense & $B\times  I \times 4\cdot d_{\mathrm{model}} $ \\
        \hspace{0.5cm} ReLU & $B\times I \times  4\cdot d_{\mathrm{model}} $ \\
        \hspace{0.5cm} Dense & $B\times I \times d_{\mathrm{model}}$ \\
        \hspace{0.5cm} Residual Connection & $B\times I \times d_{\mathrm{model}}$ \\
        \hspace{0.5cm} Normalization & $B\times I \times d_{\mathrm{model}}$ \\
        
       \hline
       Mean predictions: &  \\
    \hspace{0.5cm} Dense & $B\times d_{\mathrm{latent}}$ \\
      Variance predictions: &  \\
    \hspace{0.5cm} Dense & $B\times d_{\mathrm{latent}}$ \\
    \hline
    \hline
       
    \end{tabular}
    \caption{\textbf{Transformer encoder.} Details on the encoder architecture used for the VAE. $B$  refers to the batch size and $I$ the input dimension. For the Rydberg experiment $I=N$, where $N$ is the number of sites of the quantum system, while for the Cluster Ising model and the FKM $I=2N$. For the two latter, we set the model dimension to $d_{\mathrm{model}}=16$ and the number of heads to $N_{head}=2$, while for the former $d_{\mathrm{model}}=8$ and $N_{head}=4$. The latent space dimension was chosen to be 5 for all cases.}
    \label{tab:archi_TransfEncoder}
\end{table}

\begin{table}[h]
    \centering
    \begin{tabular}{l|c}
    \hline
    \hline
       Layer type  & Output size \\
       \hline
       Input &  $B\times I$ \\
       Embedding  & $B\times I \times d_{\mathrm{model}}$ \\
       Positional Encoding ($\sin$) & $B\times I \times d_{\mathrm{model}}$ \\
       Shift & $B\times I \times d_{\mathrm{model}}$ \\
       Transformer layer (3x):  &  \\

       \hspace{0.5cm} Normalization & $B\times I \times d_{\mathrm{model}}$ \\
       \hspace{0.5cm} Causal Self-Attention $(N_{head}=4)$ & $B\times I \times d_{\mathrm{model}}$ \\
       \hspace{0.5cm} Context Based Attention $(N_{head}=4)$ & $B\times I \times d_{\mathrm{model}}$ \\
        \hspace{0.5cm} Residual Connection & $B\times I \times d_{\mathrm{model}}$ \\
        \hspace{0.5cm} Dense & $B\times  I \times 4\cdot d_{\mathrm{model}} $ \\
        \hspace{0.5cm} ReLU & $B\times  I \times 4\cdot d_{\mathrm{model}} $ \\
        \hspace{0.5cm} Dense & $B\times I \times d_{\mathrm{model}}$ \\
        \hspace{0.5cm} Residual Connection & $B\times I \times d_{\mathrm{model}}$ \\
        \hspace{0.5cm} Normalization & $B\times I \times d_{\mathrm{model}}$ \\
        
       \hline
       Dense & $B\times I \times 2$ \\
       Log softmax & $B\times I \times 2$ \\

    \hline
    \hline
       
    \end{tabular}
    \caption{\textbf{Transformer decoder.} Details on the decoder architecture. See \cref{tab:archi_TransfEncoder} for definitions. As for the encoder, we consider two different architectures: one for the Rydberg experiment ($I=N$ and $d_{\mathrm{model}}=8$) and a slight variation for the Cluster Ising model and the FKM ($I=2N$ and $d_{\mathrm{model}}=32$). Every transformer layer performs cross-attention to the latent variables $\mathbf{z}$, embedded to dimension $B\times I \times d_{\mathrm{model}}$ by means of a linear layer.}
    \label{tab:archi_TransfDecoder}
\end{table}

The parameters of these neural networks were optimized using a gradient-based approach. To minimize the loss (presented in~\cref{eq:loss_VAE}), we used the \textit{adabelief} optimizer~\cite{Adabelief_Zhuang_2020}, with default parameters and a learning rate of 0.001 for every case except the FKM, where we used 0.0001. 

The strength of the KL regularization, controlled by the parameter $\beta$ (see \cref{eq:loss_VAE}), was set to $0.45$, $0.65$, and $0.45$ for the Rydberg atoms, cluster Ising model, and FKM, respectively. In practice, for the kind of data we are dealing with, we observe that adopting a standard value of $\beta = 0.5$ already enables a rapid identification of the number of active latent neurons while producing a meaningful latent representation. Subsequently, the performance can be further refined by fine-tuning this hyperparameter, guided by the evolution of $\log \sigma_i^2$ during training.

\section{VAE loss} \label{an:loss}

In this appendix, we provide additional clarifications regarding the loss function used throughout the numerical experiments. As introduced in the main text (see~\cref{eq:loss_VAE}), the parameters of the VAE are trained by minimizing the loss
\begin{equation}
    \mathcal{L} = -\frac{1}{M}\sum_{m=1}^{M}\mathbb{E}_q[\log p(\mathbf{x}^{m}|\mathbf{z})] - \beta \mathrm{KL}[q(\mathbf{z}|\mathbf{x}^{m})||p(\mathbf{z})],
    \label{eq:app_loss}
\end{equation}
where $M$ is the number of data samples, $p(\mathbf{z})$ denotes the prior distribution over latent variables, and $q(\mathbf{z}|\mathbf{x})$ is the variational approximation to the true posterior. The first term corresponds to the expected reconstruction log-likelihood, while the second term introduces a regularization that penalizes deviations of $q(\mathbf{z}|\mathbf{x})$ from the normal prior.

\subsection{Reconstruction loss}

In many quantum settings, like those considered in \cref{sec:results_Rybderg} and \cref{sec:cluster_ising}, we have access to samples $\mathbf{x}$ consisting of the discrete outcome of a given measurement. For example, in the case of projective measurements on a spin-$1/2$ system, a configuration can be written as
\[
\mathbf{x} = [0,1,0,0,1,1,\ldots],
\]
where the entry at position $j$ (taking values $0$ or $1$) encodes the outcome of the measurement performed on spin $j$ (up or down). More generally, for systems with $K$ possible outcomes per site, we model each local variable using a categorical distribution. This is conveniently represented via a one-hot encoding $x_i = [x_{i,1}, x_{i,2}, \ldots, x_{i,K}]$, where $x_{i,k} = 1$ if the outcome at site $i$ corresponds to class $k$, and $0$ otherwise. The associated probability mass function is $p(x_{i,k} = 1) \in [0,1]$ with the normalization condition $\sum_{k=1}^K p(x_{i,k} = 1) = 1$.

Within this framework, the reconstruction term of the VAE objective takes the form
\begin{equation}
\label{eq:reconstr_loss_discrete}
    \mathbb{E}_{\mathbf{z}\sim q(\mathbf{z}|\mathbf{x})}
    \left[\log p(\mathbf{x}|\mathbf{z})\right]
    =
    \mathbb{E}_{\mathbf{z}\sim q(\mathbf{z}|\mathbf{x})}
    \sum_{i=1}^{N}\sum_{k=1}^{K} x_{i,k} \log p(x_{i,k} = 1 |\mathbf{z}).
\end{equation}

In all discrete datasets considered in this work, namely Rydberg atom snapshots (\cref{sec:results_Rybderg}), the cluster Ising model (\cref{sec:cluster_ising}), and the $f$-fermion configurations of the Falicov--Kimball model (\cref{sec:fkm}), we have $K=2$.

To accommodate continuous data, in our case the local $d$ particle densities of the Falicov-Kimball model, we design the VAE decoder to output both the mean $\mu_d$ and the logarithm of the variance $\log \sigma^2_d$ of the density $d$, assuming a Gaussian likelihood $d \sim \mathcal{N}(\mu_d,\sigma_d^2)$.
As in the discrete setting, the reconstruction term of the objective is derived from the negative log-likelihood of the assumed output distribution. Replacing the categorical likelihood used for discrete variables by a multivariate normal distribution yields the per-sample reconstruction loss
\begin{equation}
    \mathcal{L}_{\mathrm{reconstr.}}^{d-\mathrm{particles}}
    =  \frac{1}{2}\Big[
        \log \sigma_d^2
        + \frac{(d-\mu_d)^2}{\sigma_d^2}
        + \log(2\pi)
    \Big].
\end{equation}
This formulation naturally generalizes the likelihood-based reconstruction used for discrete observables and allows the model to capture both the mean behavior and fluctuations of continuous quantum data. A hyperparameter $\alpha \in [0,1]$ is introduced to balance the two reconstruction terms
\begin{equation}
    \mathcal{L}_{\mathrm{reconstr.}}^\mathrm{FKM} = \alpha \cdot \mathcal{L}_{\mathrm{reconstr.}}^{f-\mathrm{particles}} + (1-\alpha)\cdot \mathcal{L}_{\mathrm{reconstr.}}^{d-\mathrm{particles}}
\end{equation}
where the reconstruction loss of the $f$ particles $\mathcal{L}_{\mathrm{reconstr.}}^{f-\mathrm{particles}}$ corresponds to~\cref{eq:reconstr_loss_discrete} with $K=2$. 
To produce the results shown in~\cref{sec:fkm}, we used $\alpha=0.8$.

\subsection{KL Regularization Term}
\label{app:kl}

We now provide further details on the treatment of the KL regularization term in VAE loss (see~\cref{eq:loss_VAE}). A common practice is to assume that the approximate posterior $q(\mathbf{z}|\mathbf{x})$ is Gaussian with diagonal covariance, i.e. $q(\mathbf{z}|\mathbf{x}) \sim \mathcal{N}(\boldsymbol{\mu},\boldsymbol{\Sigma})$ with $\boldsymbol{\Sigma}_{i,i}=\sigma_i^2$ and $\boldsymbol{\Sigma}_{i,j\neq i} =0$. Using a normal prior $p(z)\sim \mathcal{N}(0,1)$, the KL divergence can be written as a closed-form expression:
\begin{equation}
    \mathrm{KL}\big(q(\mathbf{z}|\mathbf{x})\,\|\,p(\mathbf{z})\big)
    = \frac{1}{2}\sum_{j}\left(\mu_j^{2} + \sigma_j^{2} - 1 - \log \sigma_j^{2}\right),
    \label{eq:app_diag_kl}
\end{equation}
where $\mu_j$ and $\sigma_j^2$ denote the mean and variance of the $j$th latent variable. This formulation is computationally efficient since it does not require estimating any terms.

\subsubsection{Comparison with other VAE objectives}

In this subsection, we compare the VAE's objective used throughout the manuscript with another common objective, namely the TCVAE~\cite{Chen_Isolating_2018}. The latter has shown a great advantage in that it decomposes the KL divergence into interpretable contributions: a mutual information term between $\mathbf{x}$ and $\mathbf{z}$, a total correlation term between the latent neurons (TC), and a dimension-wise KL penalty. In practice, all these terms are computed using statistical estimates, opposed to the usual VAE objective, which can be analytically computed (\cref{eq:app_diag_kl}). This makes their training more noisy and often more challenging.

In addition to their differences in computational cost, these two approaches impose different structural constraints on the latent representation. The analytic KL in~\cref{eq:app_diag_kl} encourages each posterior distribution $q(\mathbf{z}|\mathbf{x})$ to be individually close to the prior $p(\mathbf{z})$, resulting in latent spaces that tend to be locally smooth and exhibit limited global balancing of values across the dataset. In contrast, the TCVAE decomposition encourages the global marginal $q(\mathbf{z})$ to match the prior, while allowing individual posteriors to vary more substantially. 

\begin{figure}[h]
    \centering
    \includegraphics[width=0.85\linewidth]{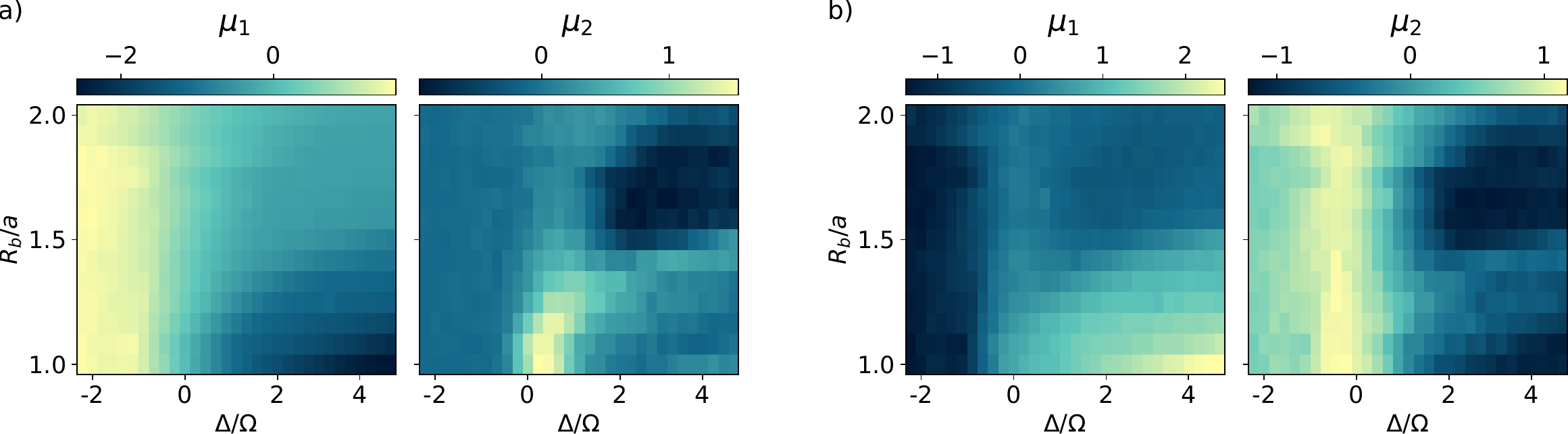}
    \caption{\textbf{Rydberg atoms: KL regularizations.}
    Representations learned by training the VAE using \textbf{a)} the analytical KL regularization employed in the main text and \textbf{b)} the total correlation (TC) decomposition of the KL divergence, as used in~\cite{de2025interpretable}.}
    \label{fig:Rydberg_TCvsAnal}
\end{figure}

We used the Rydberg dataset to compare the two KL regularization choices. The latent representations obtained from VAEs trained with each regularization scheme are shown in~\cref{fig:Rydberg_TCvsAnal}. As illustrated in the figure, the TCVAE approach produces latent variables whose values are globally balanced between positive and negative regions across the parameter space. 

In the present work, because the parameter space structures of interest are characterized by localized clusters and subtle variations, we find that the analytic KL formulation leads to latent representations better suited for resolving these features. Therefore, we adopt the closed-form KL expression in~\cref{eq:app_diag_kl} in all the numerical experiments presented.

\section{Benchmarking symbolic regression on the $J_1J_2$ model} \label{an:SM_on_J1J2}

In this appendix, we provide additional results on the symbolic regression (SR) method. We present an extensive benchmark of the method used throughout this work (SR1) against other existing approaches.


In the context of this work, the objective of SR is to find a closed-form expression $f(\mathbf{x})$ that describes the nature of an unknown cluster $\mathcal{C}$ appearing in the latent representation. In the following, we introduce three different strategies for defining such objectives.

\subsection{SR objectives}

\subsubsection*{SR1: Classification}

As shown in the main text, we can phrase the SR problem as a binary classification task: we look for an analytic expression $f(\mathbf{x})$ that separates samples belonging to the target latent cluster $\mathcal{C}$ from those that do not. Each data point $\mathbf{x}$ is assigned a label $y\in\{0,1\}$, which allows us to form a dataset $\mathcal{D}=\{(\mathbf{x}^m,y^m)\}_{m=1}^M$. For a given sample $\mathbf{x}^m$, the predicted probability is obtained via the sigmoid $\hat y^{m} = 1/(1+\mathrm{e}^{-f(\mathbf{x}^{m})})$. With these definitions, we optimize $f$ by minimizing the binary cross-entropy loss
\begin{equation}
  \mathcal{L}_{\mathrm{CE}}(y,\hat y)
  =  -\frac{1}{M}\sum_{m=1}^M\Big[
    y^{m}\log\hat y^{m}  +
    \big(1-y^{m}\big)\log\big(1-\hat y^{m}\big)
  \Big].
\end{equation}

This classification-based formulation is both simple and physically intuitive, as it directly mirrors the role of an order parameter distinguishing between phases. Nevertheless, it requires the explicit definition of \emph{in} and \emph{out} samples, and this choice may influence the structure of the learned function $f$. 

To assess the robustness of this approach, we compare it with alternative strategies proposed in the literature, which avoid the explicit selection of \emph{in} and \emph{out} samples, and therefore do not rely on sharply defined cluster boundaries.

\subsubsection*{SR2: Gradient alignment: $x-$space}

Another strategy is to frame the problem in terms of boundary geometry rather than direct classification performance~\cite{wetzel2025closed}. The key observation is that if the goal is to reproduce the decision structure of a neural network classifier $\mathrm{NN}(\mathbf{x})$, then one may instead require a symbolic expression $f(\mathbf{x})$ to match the \textit{shape} of that decision boundary. Concretely, this is achieved by aligning the input-space gradients of the symbolic model, $G_i^f = \partial f(\mathbf{x})/\partial x_i$, with those of the neural network, $G_i^{\mathrm{NN}} = \partial \mathrm{NN}(\mathbf{x})/\partial x_i$, evaluated near $\partial\mathcal{C}$.

In our setting, the encoder of the VAE naturally plays the role of such a classifier (up to constants and the final sigmoid activation), enabling a direct comparison between its gradients and those produced by candidate symbolic expressions. A simple loss function whose minimum enforces gradient alignment is the mean squared error,
\begin{equation}
    \mathcal{L}_{\text{MSE}}(G^{\mathrm{NN}},G^f)
    = \frac{1}{NM}\sum_{m=1}^M \sum_{i=1}^N \Big[ G_i^f(\mathbf{x}^{m}) - G_i^{\mathrm{NN}}(\mathbf{x}^{m}) \Big]^2.
\end{equation}
With normalized gradients, minimizing the MSE is equivalent to minimizing the cosine loss.

This formulation shifts the emphasis from label fidelity to geometric consistency, allowing symbolic regression to recover decision boundaries even when the notion of ``in'' versus ``out'' is not sharply defined. Another advantage of this approach is that it can be easily extended to match multiple clusters at the same time by just taking all boundaries. One drawback is that if the system exhibits two intertwined phases with similar boundary shapes (for example, a circle within a circle), this approach will not be able to distinguish them.

\subsubsection*{SR3: Gradient alignment: $\theta$-space}

Inspired by the previous approach, we explore an alternative objective tailored to our model and to the structure of the underlying physical problem. Consider a Hamiltonian $H(\theta_1,\theta_2,\ldots)$ depending on experimentally tunable parameters $\boldsymbol{\theta}=(\theta_1,\theta_2,\ldots)$. The samples $\mathbf{x}$ are quantum data obtained from a state related to this Hamiltonian and therefore depend implicitly on these parameters, $\mathbf{x}(\boldsymbol{\theta})$. As illustrated in~\cref{fig:proposed_pipepline}b, the latent representation $\mathrm{ENC}(\mathbf{x})\rightarrow \mu_i \sim z_i$ is visualized as a function of~$\boldsymbol{\theta}$, which implies that the cluster boundaries $\partial\mathcal{C}$ are also defined in $\boldsymbol{\theta}$-space. In this setting, matching gradients in $\mathbf{x}$-space (SR2) becomes ill-defined and over-constraining; instead, the objective should be to align gradient's direction with respect to~$\boldsymbol{\theta}$.

To do so, one needs the gradient of the encoder with respect to~$\theta_k$,
$$
G_k^{\mathrm{ENC}} := \frac{\partial\, \mathrm{ENC}[\mathbf{x}(\boldsymbol{\theta})]}{\partial \theta_k}.
$$
However, the latter is not directly accessible since the encoder takes  as input $\mathbf{x}$ rather than $\boldsymbol{\theta}$. To address this, we train a small auxiliary neural network $g$ such that $g(\boldsymbol{\theta})=\mu_i$ on a submanifold where $\mathcal{C}$ is defined. This surrogate model both approximates the mapping $\boldsymbol{\theta}\mapsto \mu_i$ and smooths out noise originating from the finite number of quantum samples. Once trained, we simply obtain the encoder gradients as
$$
G_k^{\mathrm{ENC}}=\frac{\partial g(\boldsymbol{\theta})}{\partial \theta_k}.
$$

Similarly, the symbolic function $f$ does not depend explicitly on $\boldsymbol{\theta}$. We therefore use the chain rule to project its gradient from $\mathbf{x}$-space to $\theta$-space,
$$
G_k^f := \frac{\partial f}{\partial \theta_k}
= \frac{\partial f}{\partial x}\cdot v_k,
$$
where $v_k=\partial x/\partial\theta_k$ acts as a projector. We consider two strategies to estimate $v_k$.

\paragraph*{Finite-difference projector.}
Given a sample $\mathbf{x}(\boldsymbol{\theta}^i)=\mathbf{x}^i\in\partial\mathcal{C}$ and a nearby point $\mathbf{x}^f$ reached by following the gradient direction at $\boldsymbol{\theta}^i$, we approximate
$$
v_k^{\Delta}
= \frac{x^f - x^i}{\theta_k^f - \theta_k^i}.
$$

\paragraph*{cp-based projector.}
Alternatively, we leverage the conditional probabilities provided by the VAE decoder $\mathrm{DEC}(\mathbf{z})=\mathrm{cp}$. Since $\mathrm{cp}$ are continuous, we can use another application of the chain rule:
$$
v_k^{\mathrm{cp}}
= \frac{\partial\, \mathrm{cp}}{\partial \mu_i}
\cdot
\frac{\partial \mu_i}{\partial \theta_k}.
$$
This estimator avoids finite-difference noise and imposes a more restrictive structure. In particular, it can mitigate artifacts arising from ground-state degeneracies and other non-smooth variations in the data manifold. One drawback is that, since it relies on the cp and its chosen ordering, the interpretability of the final symbolic expression may be deteriorated.

\subsection{Results} \label{an:SM_on_J1J2_results}

We now test the symbolic approaches introduced in the previous section. As a controlled setting, we study a well-known quantum spin system: the $J_1J_2$ model on a $3\times 3$ square lattice with open boundary conditions~\cite{jin2013phase}.  
Fixing $J_1 = 1$, the Hamiltonian reads
\begin{equation}
    H_{J_1J_2}
    = 
    \sum_{\langle i,j\rangle} \sigma_i^z \sigma_j^z
    +
    J_2 \sum_{\langle\langle i,j \rangle\rangle} \sigma_i^z \sigma_j^z
    +
    h\sum_i \sigma_i^x ,
\end{equation}
where $\langle \cdot\rangle$ and $\langle\langle \cdot\rangle\rangle$ denote nearest-neighbor (NN) and next-nearest-neighbor (NNN) pairs, respectively.  
We treat the two experimental tuning parameters as $\theta_1 = J_2 \in [0,1.5]$ and $\theta_2 = h \in [0.1,2]$.

\begin{figure}
    \centering
    \includegraphics[width=13.5cm]{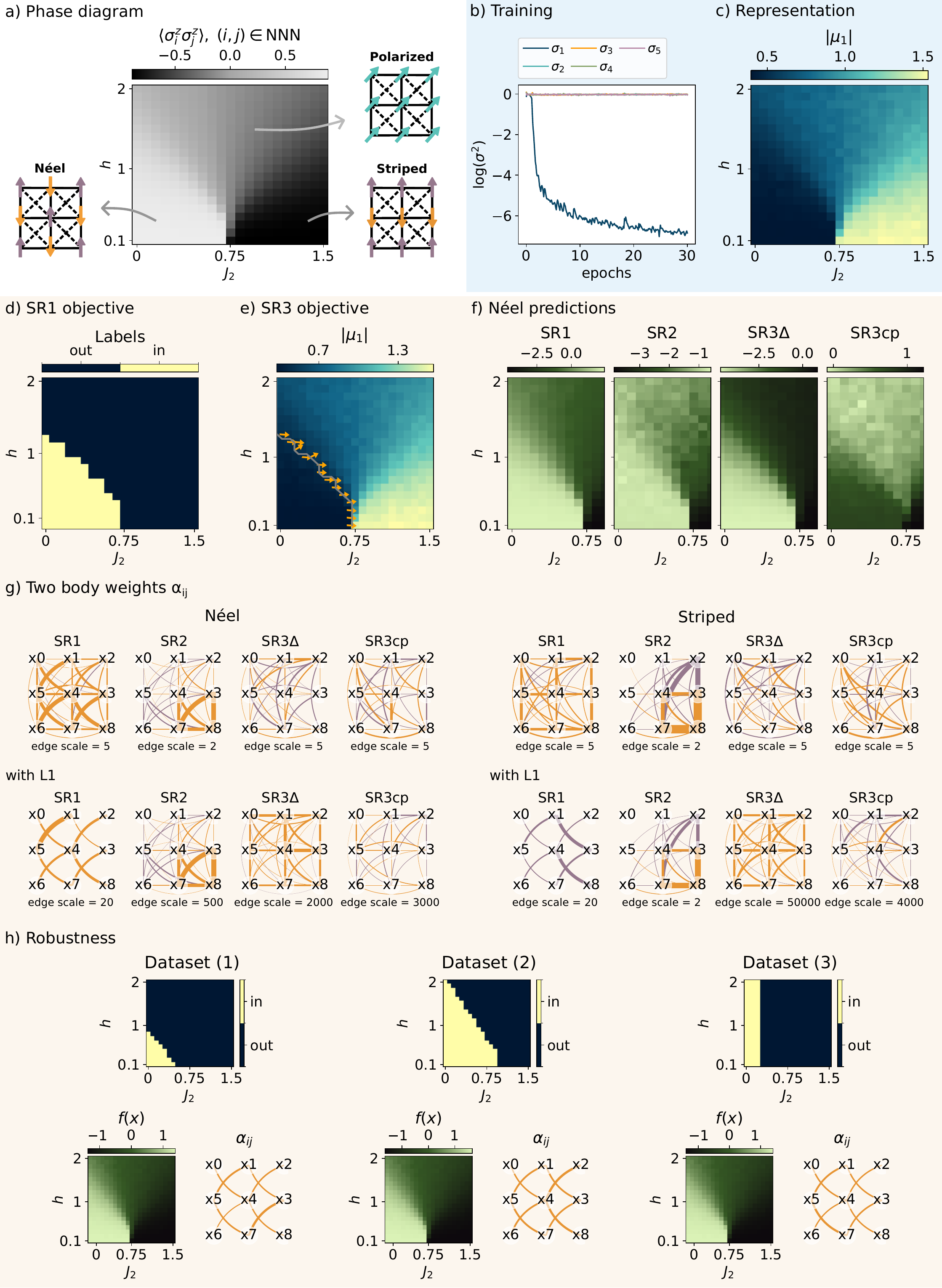}
    \caption{\textbf{$J_1J_2$ model.} 
    \textbf{a)} Phase diagram visualized using the next–nearest-neighbor (NNN) correlator.  
    \textbf{b)} Training of the VAE seen through the log of the variances $\log(\sigma_i^2)$ output by the encoder. 
    \textbf{c)} Latent representation learned by the VAE.
    \textbf{d)} Classification objective of the first symbolic regression approach (SR1).
    \textbf{e)} Gradient alignment objective of the third symbolic regression methods (SR3).
    \textbf{f)} Predictions of the symbolic function $f(\mathbf{x})$ obtained after training with the various SR objectives to characterize the Néel phase.
    \textbf{g)} Learned two-body correlator weights $\alpha_{ij}$ of $f(\mathbf{x})$. Links are colored according to the sign of $\alpha_{ij}$, orange being positive and purple negative. Their thickness scales with $|\alpha_{ij}|$. Results with an additional $\text{L1}$ regularization are presented in the bottom rows.
    \textbf{h)} Prediction $f(\mathbf{x})$ and its two-body correlator weights $\alpha_{ij}$ learned with SR1 for different classification datasets. Results are presented with the additional $\text{L1}$ regularization.
}
    \label{fig:J1J2}
\end{figure}

Across this parameter space, the system contains three distinct phases (see~\cref{fig:J1J2}a). For small $J_2$ and small $h$, the nearest-neighbor (NN) interactions dominate, stabilizing a Néel-ordered phase characterized by antiferromagnetically aligned NN spins. For large $J_2$ and small $h$, the next-nearest-neighbor (NNN) couplings prevail, giving rise to a striped phase in which NN spins align while NNN spins are antiferromagnetically aligned. Finally, for sufficiently large transverse field $h$, the spins polarize along the $x$-direction, forming the polarized phase~\cite{sadrzadeh2018phase}.

Given the simplicity of the data, we build the VAE encoder using a convolutional neural network. The details of its architecture are presented in \cref{tab:archi_convEncoder}. For the decoder, a standard feedforward neural network (FFNN) was used. Autoregression is achieved by simply imposing triangular weights. For the numerical simulations, we used 3 hidden layers with 36 hidden neurons and SeLU activation functions~\cite{Self-Normalizing_Klambauer_2017}.

\begin{table}[h]
    \centering
    \begin{tabular}{l|c}
    \hline
    \hline
       Layer type  & Output size \\
       \hline
       Input &  $B\times L \times L \times 1$ \\
       Circular Conv 2D ($k_s=3\times3$)  & $B\times L \times L \times 32$ \\
       ReLu  & $B\times L \times L \times 32$ \\
       Circular Conv 2D ($k_s=3\times3$)  & $B\times L \times L \times 32$ \\
       ReLu  & $B\times L \times L \times 32$ \\
       Global Average Pooling  & $B\times 32$ \\
       \hline
       Mean predictions: &  \\
    \hspace{0.5cm} Dense & $B\times 32$ \\
    \hspace{0.5cm} ReLu & $B\times 32$ \\
    \hspace{0.5cm} Dense & $B\times 5$ \\
      Variance predictions: &  \\
    \hspace{0.5cm} Dense & $B\times 32$ \\
    \hspace{0.5cm} ReLU & $B\times 32$ \\
    \hspace{0.5cm} Dense & $B\times 5$ \\
    \hline
    \hline
       
    \end{tabular}
    \caption{\textbf{Convolutional encoder.} Details on the encoder architecture used for the $J_1J_2$ model. $B$ is the batch size, $L=\sqrt{N}$ the size of the square lattice and $k_s$ the kernel size. The latent space dimension was chosen to be 5. }
    \label{tab:archi_convEncoder}
\end{table}

We trained the VAE on an ensemble of unlabeled spin configurations coming from various points across the $(J_2,h)$ parameter space. During training, we observe that a single latent neuron remains active while all others collapse to the prior (see~\cref{fig:J1J2}b). This behavior is consistent with the full phase diagram being effectively parameterized by a single degree of freedom, such as the next–nearest-neighbor (NNN) correlator (see~\cref{fig:J1J2}a). The active latent variable encodes a quantity related to the magnetization, and due to the $\mathbb{Z}_2$ degeneracy, it averages to zero. Thus in order to visualize the latent representation, we show the average of its absolute value across the parameter space in~\cref{fig:J1J2}c.

\subsubsection{Quantitative benchmark}

To benchmark the three symbolic regression methods (SR) under controlled and comparable conditions, we restrict the symbolic expression to the quadratic form
\begin{equation} \label{eq:ansatz_fx}
    f(\mathbf{x}) = \sum_{i,j>i} \alpha_{i,j}\; x_i x_j ,
\end{equation}
where $(i,j)$ runs over all spin pairs and the coefficients $\alpha_{i,j}$ are the trainable parameters. For each SR, the coefficients $\alpha_{i,j}$ are optimized with the L-BFGS-B routine from \texttt{scipy}, minimizing the corresponding loss function. In the case of characterizing the Néel phase, the classification objective of SR1 as well as the gradient alignment objective of SR3 can be visualized in~\cref{fig:J1J2}d and~\ref{fig:J1J2}e, respectively.

\begin{table*}[b]
\centering
\caption{\textbf{Comparison of symbolic methods.} 
Performance of the symbolic regression methods (SR) in identifying the Néel, striped, and polarized phases. 
For each task, we report the Spearman correlation between the learned symbolic function $f(\mathbf{x})$ and the corresponding order parameter as well as the classification AUC scores. The best results are presented in bold.
}
\vspace{0.5em}
\setlength{\tabcolsep}{7pt}
\renewcommand{\arraystretch}{1.15}
\resizebox{1.\textwidth}{!}{
\begin{tabular}{ll
                cc
                cc
                cc}
\toprule
\multirow{2}{*}{Approach} & \multirow{2}{*}{Variant}
    & \multicolumn{2}{c}{\textbf{Néel phase}}
    & \multicolumn{2}{c}{\textbf{Striped phase}}
    & \multicolumn{2}{c}{\textbf{Polarized phase}} \\
\cmidrule(lr){3-4} \cmidrule(lr){5-6} \cmidrule(lr){7-8}

 & 
 & $r_s(f(\mathbf{x}),\mathrm{NN})$
 & AUC
 & $r_s(f(\mathbf{x}),\mathrm{NNN})$
 & AUC
 & $r_s(f(\mathbf{x}),\mathrm{|NNN|})$
 & AUC \\

\midrule

SR1: in vs out cls. 
& -- 
 & 0.977 & 0.999 
 & 0.986 & \textbf{1.000} 
 & 0.944 & 0.994 \\ 

\midrule

SR2: grad alignment (x-space) & --
 & 0.868 & 0.980 
 & 0.912 & 0.985 
 & 0.003 & 0.531 \\ 

\midrule

SR3 (proposed):
 & vk from $\Delta x$
 & \textbf{0.995} & \textbf{1.000} 
 & \textbf{0.996} & \textbf{1.000} 
 & \textbf{0.980} & \textbf{0.998} \\ 
 grad alignment ($\theta$-space)
 & vk from cp
 & 0.888 & 0.997 
 & 0.854 & 0.932 
 & 0.905 & 0.970 \\ 

\bottomrule
\end{tabular}
}\label{tab:SM_results}
\end{table*}

We benchmark the different symbolic regression (SR) strategies by evaluating the learned expressions $f$ on three tasks: characterizing the Néel phase, characterizing the striped phase, and characterizing the polarized phase. For each task, we compute the Spearman correlation between $f$ and the corresponding physical order parameter: i) the nearest-neighbor (NN) correlator for the Néel phase; ii) the next-nearest-neighbor (NNN) correlator for the striped phase; iii) and the absolute value of the NNN correlator for the polarized phase. In addition, we report the area under the curve (AUC) to quantify how well $f$ separates the relevant phases. Given a set of quantum data $\mathbf{x}$ labeled as ``in-phase'' or ``out-of-phase'', the scalar output $f$ can be interpreted as a ranking score. The AUC is then equal to the probability that a randomly chosen in-phase configuration receives a larger score than a randomly chosen out-of-phase configuration~\cite{hanley1982meaning},
\begin{equation}
    \mathrm{AUC}
    \;=\;
    \Pr\!\big(f(\mathbf{x}^{\mathrm{in}}) > f(\mathbf{x}^{\mathrm{out}})\big).
\end{equation}
An AUC of $1$ indicates perfect separability of the two score distributions, while an AUC of $0.5$ corresponds to random guessing. We restrict the evaluations of both metrics to the left (resp. right) half of the parameter space for the Néel (resp. striped) phase.

For the Néel phase, the prediction landscape $f(\mathbf{x})$ over the left half of the parameter space is shown in~\cref{fig:J1J2}f. With the exception of SR2, all methods enable to clearly distinguish the Néel phase. The predictions obtained with the SR3cp objective exhibit a higher level of noise, which we attribute to the use of the decoder conditional probabilities. Since these probabilities are learned from the data, they inherit part of its noise, thereby affecting the symbolic fit. In contrast, SR1 and SR3$\Delta$ yield smoother decision landscapes.

These qualitative observations are confirmed by the quantitative metrics reported in~\cref{tab:SM_results}, where the best scores are highlighted in bold. Across all three tasks and both evaluation measures, SR3$\Delta$ achieves the highest overall scores, with SR1 performing comparably and, in some cases, matching its performance. The slightly lower scores of SR1 can be attributed to its reliance on a threshold criterion (e.g.\ $|\mu_1| < \delta$) to define phase boundaries. Modifying $\delta$ directly shifts the classification boundary and thus affects its objective. By contrast, SR3 are less sensitive to such shifts in boundary location since the gradients keep the same orientations.

\subsubsection{Interpretability benchmark}

Beyond predictive performance, interpretability is a key criterion. In this context, we examine whether the learned coefficients $\alpha_{ij}$ reveal a physically meaningful structure. The coefficients obtained for the Néel and striped classifiers are shown in~\cref{fig:J1J2}g, where links are colored according to the sign of $\alpha_{ij}$ and their thickness is proportional to $|\alpha_{ij}|$. All symbolic expressions involve many two-body couplings. However, the coefficients obtained with the SR1 objective display a more structured pattern. In contrast, those resulting from the SR2 objective are strongly localized on a small subset of spins, while the coefficients learned with SR3 do not exhibit a clear spatial pattern.

In the bottom row of~\cref{fig:J1J2}g, we also show the coefficients obtained when each objective is optimized with an additional $\text{L1}$ regularization term. This term corresponds to the mean absolute value of the parameters: $\text{L1} = \sum_{i,j>i}|\alpha_{i,j}|$. It promotes sparsity and can enhance interpretability. When applied to SR2 and SR3, the $\text{L1}$ penalty mainly reduces the magnitude of the coefficients without significantly sparsifying their structure. By contrast, applying $\text{L1}$ regularization to SR1 produces a markedly sparser set of coefficients, thereby improving interpretability. In this case, the recovered structure closely matches the true order parameter, namely the next-nearest-neighbor (NNN) correlator.

Interestingly, without $\text{L1}$ regularization, the coefficients obtained with SR1 for the striped phase predominantly highlight nearest-neighbor (NN) links. However, NN correlations alone are insufficient to isolate the striped phase. As a result, the prediction is also driven by many weaker couplings that cannot be neglected. Introducing $\text{L1}$ regularization guides the optimization toward the correct sparse structure associated with the relevant NNN correlator.

\subsubsection{Robustness against misclassification} \label{an:J1J2_res_rob}

As a final test, we assess the robustness of the SR classification scheme (SR1) in identifying the Néel phase when the 'in' and 'out' labels are perturbed. Specifically, we shift the boundaries of the ground-truth classification (see left panel of~\cref{fig:J1J2}d), thereby introducing controlled mislabelling between the two classes. The resulting classification tasks are displayed in the first row of~\cref{fig:J1J2}h and the amount of mislabelling is summarized in~\cref{tab:SR1_robu}. For each case, we also report the Spearman correlation between the learned function $f(\textbf{x})$ and the nearest-neighbor (NN) correlator, as well as the corresponding AUC score. Results are presented both without and with the additional $\text{L1}$ regularization term. Overall, SR1 demonstrates remarkable robustness to shifts in the location of the target cluster. Without $\text{L1}$ regularization, shifts even lead to slight improvements in both the Spearman correlation and AUC, which can be attributed to slightly less overfitting. Incorporating $\text{L1}$ regularization further enhances both performance and stability. The reported results are the mean of 10 runs with various initial weights. With standard deviations always below 0.001, this also underscores the robustness of the method against initial conditions. 

The predicted function $f(\textbf{x})$ consistently highlights the Néel phase across all perturbed datasets considered (see second row of~\cref{fig:J1J2}h). In addition, the structure of the learned two-body coefficients remains intact, still very close to the NNN correlator.

\begin{table*}[t]
\centering
\caption{\textbf{Robustness of SR1.} 
Performance of the SR classification scheme (SR1) in identifying the Néel in the presence of a noise in the 'in' and 'out' labels in its dataset.
For each case, we report the Spearman correlation between the learned symbolic function $f(\mathbf{x})$ and the corresponding order parameter as well as the classification AUC scores. 
}
\vspace{0.5em}
\setlength{\tabcolsep}{7pt}
{
\begin{tabular}{ll
                c
                c
                c}
\toprule
    &
    & Dataset (1)
    & Dataset (2) 
    & Dataset (3) \\

\midrule

\vspace{0.1cm}

True 'in' &
 & 100\% 
 & 54\% 
 & 54\%  \\

True 'out' &
 & 87\% 
 & 100\% 
 & 91\%  \\

\midrule

\vspace{0.1cm}

\multirow{2}{*}{$r_s(f(\mathbf{x}),\mathrm{NN})$}
 & without L1
 & 0.969 & 0.990 & 0.980 \\ 

 & with L1
 & 0.990 & 0.990 & 0.990 \\ 

\midrule

\vspace{0.1cm}

\multirow{2}{*}{AUC}
 & without L1
 & 0.999 & 1.000 & 0.999 \\ 

 & with L1
 & 1.000 & 1.000 & 1.000 \\

\bottomrule
\end{tabular}
}\label{tab:SR1_robu}
\end{table*}

\subsubsection{Genetic search with broader search spaces}
\label{sec:SR_broader}

While the two-body correlator ansatz (see~\cref{eq:ansatz_fx}) provides a controlled and interpretable setting to benchmark the different objectives, such a simple ansatz may not always be adequate for more complex systems. In this case, genetic-algorithm-based symbolic regression offers a flexible alternative, enabling the exploration of large spaces of candidate expressions. To illustrate this approach, we apply it to the task of classifying Néel configurations. Starting from the following set of elementary operations: $+$, $-$, $/$, $\cdot$, $\sin$, $\cos$, $\tanh$, $\exp$ and $\log$, the algorithm can construct arbitrary functions up to a prescribed level of complexity.

The resulting search space includes not only two-body terms, but also one-body or higher-order contributions and complex trigonometric or exponential terms. Despite this increased search space, the symbolic regression procedure converges to the following expression,
\begin{equation*}
    f(\mathbf{x}) = 0.64\,(x_0 x_4 + x_1 x_3 + x_1 x_5 + x_2 x_4 + x_3 x_7 + x_4 x_6 + x_4 x_8 + x_5 x_7) - 2.44,
\end{equation*}
which corresponds, up to an additive constant, to the next-nearest-neighbor correlator. This result demonstrates that genetic-algorithm-based symbolic regression can successfully recover relevant order parameters even when exploring broad and weakly constrained function spaces. Nevertheless, we note that the present setting involves only $9$ spins, such that the dimensionality of the input space remains relatively small. In scenarios involving larger systems, we believe that incorporating prior knowledge, as for instance that offered by the output probabilities of the VAE (see \cref{sec:results_Rybderg} and \cref{sec:fkm}) to constrain the search space of candidate expressions can be beneficial, both by improving convergence and by guiding the discovery toward physically meaningful solutions.

\section{Details on the quantum systems}\label{an:QSyst_details}

In this appendix, we provide additional details on three quantum systems taken into account in the main text.

\subsection{Rydberg atoms}\label{an:Rydberg_details}

The experimental setup consists of a two-dimensional square lattice with $N=13\times 13$ atoms, where each atom resides in either its electronic ground state $\ket{g}$ or an excited Rydberg state $\ket{r}$. The latter is achieved via laser-driven excitation of an outer electron to a high-lying energy level, giving rise to strong, long-range interactions between excited atoms. The Rydberg array under study is characterized by the Hamiltonian

\begin{equation}
    H_{\mathrm{Rydberg}} =  \sum_{i<j} \frac{C_6}{||\mathbf{r_i}-\mathbf{r_j}||^6} \hat{n}_i\hat{n}_j - \Delta\sum_i \hat{n}_i  - \frac{\Omega}{2} \sum_i \sigma_i^x
\end{equation}
with
\begin{equation}
    \hat{n}_i = \frac{1}{2}(\sigma_i^z+\mathcal{I}) \hspace{0.5cm} \mathrm{and} \hspace{0.5cm} C_6 = \Omega \Big(\frac{R_b}{a}\Big)^6,
\end{equation}
where $\sigma_i^x= \ket{g}_i\bra{r}_i + \ket{r}_i\bra{g}_i$, $\sigma_i^z= \ket{r}_i\bra{r}_i - \ket{g}_i\bra{g}_i$, $\Delta$ is the detuning from resonance, $\Omega$ the Rabi frequency, $a$ the lattice length scale and $R_b$ the Rydberg blockade radius. A key feature of Rydberg atoms is that, when a given atom is in the excited state $\ket{r}$, the van der Waals potential effectively prevents the excitation of any other atom within a radius $R_b$, an effect commonly referred to as the ``Rydberg blockade''~\cite{lukin2001dipole,urban2009observation}. During the experiment, two different parameters were tuned: $R_b/a$, controlling the strength of the blockade, and $\Delta/\Omega$, encoding the relative strength of a longitudinal field. In particular, to create the snapshots, the neutral atoms went through a coherent quantum evolution for different values of blockade strength $R_b/a$ and relative field strength $\Delta/\Omega$. 
Then, a projective readout is performed in order to determine which atoms were not in the excited Rydberg state $\ket{r}$~\cite{Machine_Miles_2023, ebadi_quantum_2021}.

We had access to $250$ snapshots taken for each set of parameters $(R_b/a,\Delta/\Omega)$. The parameters are
\begin{equation}
    R_b/a \in \{1.01, 1.05, 1.13, 1.23, 1.3 , 1.39, 1.46, 1.56, 1.65, 1.71, 1.81, 1.89, 1.97\}
\end{equation}
and 31 values of $\Delta/\Omega \in [-2.33,4.65]$, linearly spaced. Various ordering choices are possible for modelling conditional probabilities in 2D systems. For this system, a snake ordering has been used~\cite{RydbergGPT_Fitzek_2024}. Determining the optimal ordering is an interesting task that goes beyond the scope of this work.

\subsection{Cluster Ising model} \label{an:clusterIsing_details}

We consider here the one-dimensional cluster–Ising Hamiltonian on an open chain of length $N=15$~\cite{smacchia2011statistical, verresen2017one},
\begin{equation}
\label{eq:cluster-ising}
H_{\mathrm{cluster}} =  -\sum_{i=1}^{N} \sigma^z_{i-1} \sigma^x_i \sigma^z_{i+1}
   - h_1 \sum_{i=1}^{N} \sigma^x_i  -h_2 \sum_{i=1}^{N-1}\sigma_i^x\sigma_{i+1}^x
\end{equation}
where $\sigma^x_i,\sigma^y_i,\sigma^z_i$ are Pauli operators acting on site $i$. We consider open boundary conditions defined by $\sigma^z_0 = \sigma^z_{N+1} = \mathbb{1}$. In our numerical experiments we sweep the
transverse field strength $h_1\in[0.05,\,1.20]$ and the Ising coupling $h_2\in[-1.5,\,1.5]$ to probe the competition between the cluster term, the transverse-field (paramagnetic) term, and nearest-neighbor Ising correlations. We used 24 (resp. 30) linearly spaced values for $h_1$ (resp. $h_2$). To compose the dataset, 2000 shadows were created for each set of values of $h_1$ and $h_2$.

As displayed in~\cref{fig:clusterIsing}a, the model contains three canonical regimes~\cite{khosrojerdi2025unsupervised, Arnold_Mapping_2024, herrmann_Realizing_2022, Zapletal_Error_2024, Sadoune_Unsupervised_2023}. First, for high $h_2$, the nearest-neighbor term dominates and the ground state is antiferromagnetic in the $x$-direction. Then, for low values of $h_2$ and sufficiently high $h_1$, the spins align in the $\textit{x}$ direction, leading to a paramagnetic regime. Finally, for intermediate values of $h_2$, there is a cluster phase protected by a $\mathbb{Z}_2\times\mathbb{Z}_2$ symmetry and governed by the stabilizer-like three-spin interaction $\sigma_{i-1}^z\sigma_i^x \sigma_{i+1}^z$, which on open chains yields near-degenerate edge modes and nonlocal string order~\cite{briegel2001persistent}.  
 

On an open chain the SPT phase is most clearly identified via a nonlocal string-order parameter,
\begin{equation}\label{eq:string-order}
\mathcal{O}_{\rm string}(i,j) = \big\langle \sigma_i^z \Big(\prod_{k=i+1}^{j-1} \sigma_k^x\Big) \sigma_j ^z\big\rangle,
\end{equation}
with endpoints chosen near the chain boundaries (e.g., $i=1$, $j=N$) to highlight edge correlations.

\subsection{Falicov-Kimball model} \label{an:FKM_details}

Finally, we consider the spinless Falicov-Kimball model (FKM) at half filling on a $L\times L$ square lattice, described by the Hamiltonian
\begin{equation}
    H =\; -t \sum_{\langle i,j\rangle} \big( d_i^\dagger d_j + d_j^\dagger d_i \big) + U \sum_i \left( f_i^\dagger f_i - \tfrac{1}{2} \right)\left( d_i^\dagger d_i - \tfrac{1}{2} \right).
\end{equation}
Here, $t$ denotes the nearest-neighbor hopping amplitude of the itinerant $d$ fermions and is set to 1. The second term describes a local repulsive interaction of strength $U$ between the localized (heavy) $f$ fermions and the itinerant (light) $d$ fermions occupying the same lattice site. The shifts by $-1/2$ enforce particle-hole symmetry and guarantee half filling.

The phase diagram is organized around two broad regimes: an \emph{ordered phase} (OP) and a \emph{disordered phase} (DP). In the OP, both the localized $f$ particles and the itinerant $d$ particles develop charge-density-wave (CDW) order. Outside this region, the $f$-particle configurations become disordered. Nevertheless, the behavior of the $d$ particles allows the disordered regime to be further subdivided into several distinct phases.

At strong interaction strength $U$, the system enters a gapped regime characterized by a Mott-insulating (MI) state. In this phase, the $d$ particles are tied to the $f$-particle occupation, such that $\langle n_{d,i} \rangle \approx 0$ when $n_{f,i}=1$ and $\langle n_{d,i} \rangle \approx 1$ when $n_{f,i}=0$. At weaker interactions, the disordered phase becomes gapless and can be further separated into a weakly localized regime with nearly uniform $d$-particle densities and an Anderson-localized regime with inhomogeneity. Finally, in the non-interacting limit $U=0$, the model reduces to a free Fermi gas (FG)~\cite{antipov2016interaction, frk2025unsupervised}.

The quantum data analyzed in this section come from a system size of $N=4\times4$. One part of the data corresponds to the configuration of the localized $f$ particles, with binary entries $0$ and $1$ indicating unoccupied and occupied sites, respectively. The remaining part corresponds to the local densities of the itinerant $d$ particles and takes continuous values in the interval $[0,1]$. The ordering used to model the conditional probabilities was simply done per row and per site. For each site, we first take the $f$ particle and then the $d$ particle. We treat the interaction strength $U\in[0,12]$ and the temperature $T\in[0.005,0.3]$ as tuning parameters. We used 60 (resp. 49) linearly spaced values for $U$ (resp. $T$). To compose the dataset, 1000 configurations were created for each set of values of $T$ and $U$.  The data was created from Monte Carlo simulations; further details can be found in~\cite{frk2025unsupervised, Zonda2024}.

\section{Robustness of the symbolic expressions} \label{an:robustness_symb_expr}

In this appendix, we present additional analysis on the robustness of the symbolic expressions found in the main text.

For the Rydberg array (presented in \cref{sec:results_Rybderg}), the corner order parameter $f(x) = x_1x_4+\alpha x_2x_3$ has been found in 9 of 10 symbolic searches with different initial random seeds. Out of those runs, $\alpha$ has a median of 2.0 and interquartile range (1.0, 2.1). We report medians and interquartile ranges rather than means and standard deviations since symbolic searches are known to occasionally converge to distinct optimization basins \cite{kronberger2024inefficiency}. These behave as outliers rather than symmetric noise, biasing the mean while keeping the median robust. 

For the cluster Ising model (presented in \cref{sec:cluster_ising}), the order parameter is obtained in two stages. First, the two-body weights $\alpha_{ij}$ are optimized (10 restarts with different initializations). The run reaching the lowest loss yields the weights reported in the main text. Second, these weights are compressed into a closed form by 10 independent genetic searches. Every search recovered the functional form $g(i,j) = -a_1 + a_2/(r_{ij} - a_3)$, with medians (interquartile ranges) $a_1 = 0.2021 \; (0.2018, 0.2026)$, $a_2 = 0.0700 \; (0.06966, 0.07040)$, and $a_3 = 0.9465 \; (0.9463, 0.9468)$, in close agreement with the main-text values $(0.20, 0.07, 0.95)$. 
Out of the final family of symbolic expressions, our selection criterion is PySR's standard score (maximal accuracy per unit complexity).

We further note that robustness against label misclassification of the SR trainset is demonstrated in \cref{an:J1J2_res_rob}.

\section{Classical Shadows} \label{an:classical_shadows}

This appendix provides more details on the classical shadow formalism and explains how to calculate expectation values from them. Classical shadows are a framework developed for efficiently representing quantum states based on randomized measurements~\cite{huang2022learning}. The key idea is to replace full quantum state tomography, which scales exponentially with system size, by a compact randomized data structure that retains sufficient information to estimate a wide range of physical observables.

Classical shadows address a central bottleneck in quantum experiments: the difficulty of extracting useful information from large-scale quantum states. Instead of reconstructing the full state, one can directly estimate physically relevant quantities such as correlation functions, fidelities, or entanglement measures. This makes shadows particularly well suited for current quantum hardware, where repeated measurements are costly and classical post-processing must remain efficient. In this work, we treat classical shadows as a natural form of stochastic quantum data, akin to projective measurement snapshots but with a richer structure.

We employ classical shadow tomography based on random local Pauli measurements~\cite{huang2022learning}. Consider an N-qubit quantum state $\rho$. For each shot, every qubit is measured in a randomly chosen Pauli basis $P_i \in \{X,Y,Z\}$ with equal probability. The full measurement basis is then a string $P=(P_1,...,P_N)$, and the corresponding projective measurement yields outcomes $b=(b_1,...,b_N)$, where each $b_i=\pm 1$. The pair $(P,b)$ constitutes a single classical shadow of the state. Repeating this process generates an ensemble of shadows distributed according to
\begin{align}
    p(P,b|\rho) & = p(P)p(b|P,\rho), \\
    p(P) & = 3^{-N}, \\
    p(b|P,\rho) & = \mathrm{Tr}[\rho \bigotimes_i \frac{1}{2}(1+b_i P_i)]
\end{align}

Each measurement record can be inverted via a simple channel map to obtain an unbiased single-shot estimator of $\rho$:

\begin{equation}
    \hat{\rho}(P,b) = \bigotimes_i \frac{1+3b_i P_i}{2}
\end{equation}

such that the ensemble average recovers the full state,

\begin{equation}
    \mathbb{E}_{(P,b)}[\hat{\rho}(P,b)] = \rho
\end{equation}

While full state reconstruction remains exponentially costly, classical shadows circumvent this by allowing efficient estimation of many observables without reconstructing $\rho$ explicitly.

In particular, physical observables are extracted directly from the shadow ensemble using the method of moments (MoM). Given an observable $O$, one constructs an unbiased estimator

\begin{equation}
     \hat{o}(P,b) = \mathrm{Tr}[O\hat{\rho}(P,b)].
\end{equation}

In practice, the dataset of $M$ shadow samples is partitioned into $K$ groups, the average within each group is computed, and the median of these group means is taken as the final estimate. This MoM procedure significantly suppresses the impact of heavy-tailed fluctuations and improves the stability of shadow-based estimators.

\section{Additional results on the cluster Ising model} \label{an:cluster_Ising_add_res}

\begin{figure}[h]
    \centering
    \includegraphics[width=1.\linewidth]{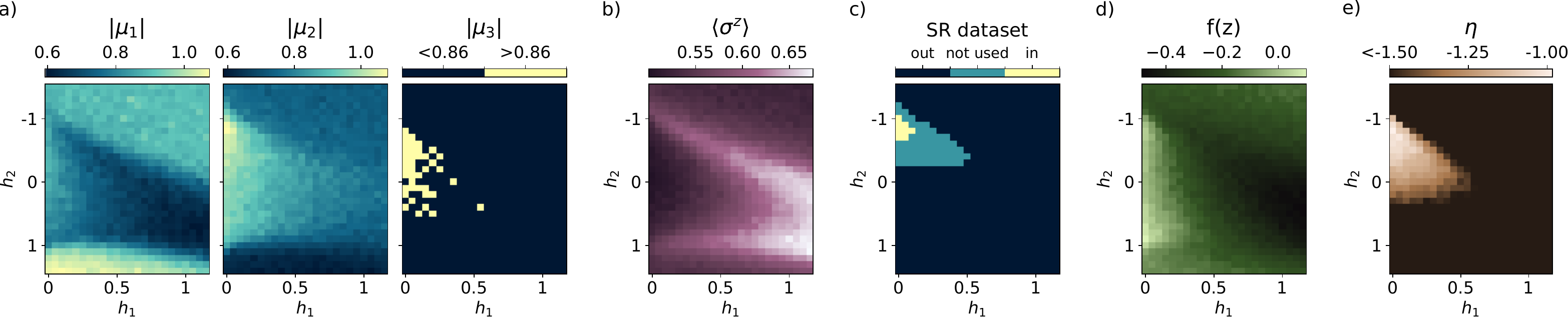}
    \caption{\textbf{Cluster Ising model: additional results.}
    \textbf{a)} Average absolute values of three active latent variables across the parameter space.
    \textbf{b)} Longitudinal $Z$ magnetization across the parameter space.
    \textbf{c)} Dataset used for the symbolic regression (SR).
    \textbf{d)} Prediction of the two-body correlator obtained when restricting the input to $Z$-basis measurements.
    \textbf{e)} Exponent $\eta$ extracted from a power-law fit of the $X$-bubbles distribution $P(s)\sim s^\eta$.}
    \label{fig:clusterIsing_appendix}
\end{figure}

In this appendix, we provide complementary results for the cluster Ising model to those presented in \cref{sec:cluster_ising}. First, in~\cref{fig:clusterIsing_appendix}a, we show the average absolute values of the first two active latent variables, $|\mu_1|$ and $|\mu_2|$, across the parameter space. This reveals complementary information encoded in the latent representation. In particular, $|\mu_1|$ closely follows the behavior of the $Z$ magnetization, shown in~\cref{fig:clusterIsing_appendix}b. Then $|\mu_2|$ further highlights the presence of the additional cluster. We also show the region where $|\mu_3|>0.86$, which again highlights the presence of the additional cluster.

To verify that the additional cluster is not an artifact of the chosen $\beta = 0.65$, we trained the VAE with $\beta = 0.75$. The representation then collapses to two active neurons, at the cost of a 120\% increase in reconstruction loss and increases of 64.6\% and 4.06\% in the errors on the reconstructed energy and three-body stabilizer, respectively. The learned representation encoded in the 2 active latent neurons is qualitatively similar to the three-active-neuron model, up to the now-inactive third neuron. In particular, the additional cluster remains encoded in the absolute value of the second latent neuron, showing that the discovery is not an artifact of the chosen $\beta$.

The classification dataset constructed to learn a symbolic function characterizing this previously unknown cluster is shown in~\cref{fig:clusterIsing_appendix}c. The resulting symbolic prediction obtained when restricting the input to $Z$-basis measurements is presented in~\cref{fig:clusterIsing_appendix}d. In this case, the symbolic model fails to clearly separate the data, demonstrating that $Z$ measurements alone are insufficient to fully characterize this regime.

At phase transitions, observables of quantum systems typically exhibit scale-free properties, showcased by power-laws on particular scaling factors~\cite{vojta2003quantum}. Therefore, to further characterize the decay of the $X$-bubbles distribution, we fit its distribution using the form $P(s)\sim s^\eta$. The resulting exponent $\eta$, displayed in~\cref{fig:clusterIsing_appendix}e, reaches its highest values within the additional cluster. We report in~\cref{tab:Bubble_fit_quality} the quality of the fits. For comparison, we also include the results using an exponential form $P(s)\sim e^{-s/\xi}$. The reported metrics are evaluated over the relevant region of the parameter space (where $\eta > -1.5$). The exponential fits perform significantly worse, thereby justifying the use of a power-law form, which provides a good description of the distribution. It is worth noting that $\eta$ reaches its maximum at the same location as $f(\textbf{x})$, which is at $(h_1,h_2)=(0.05, -0.7)$. This observation supports the idea that this unknown regime is linked to the algebraic decay of the $X$-bubble distribution and is caused by the phase transition and finite-size effects.

\begin{table*}[t]
\centering
\begin{tabular}{l@{\hspace{1cm}}
                S[table-format=1.3]
                S[table-format=1.3]
                S[table-format=1.3]
                S[table-format=1.3]}
\toprule
\multirow{2}{*}{\textbf{Fit form}} & \multicolumn{2}{c}{$R^2$ (higher is better)}
& \multicolumn{2}{c}{RMSE (lower is better)} \\
\cmidrule(lr){2-3} \cmidrule(lr){4-5}

& {Min.} & {Max.}
& {Min.} & {Max.} \\

\midrule

\vspace{0.1cm}

Power law
& 0.989 & 0.999
& 0.004 & 0.015 \\

Exponential
& 0.571 & 0.778
& 0.059 & 0.103 \\

\bottomrule
\end{tabular}
\caption{\textbf{Cluster Ising model: fits of the $X$-bubbles distribution.}
Summary of the power law and exponential fits to the $X$-bubble size distribution $P(s)$ in the cluster Ising model. Reported are the minimum and maximum values of the coefficient of determination $R^2$ (higher is better) and the root-mean-square error (RMSE, lower is better) across the parameter region of interest (with $\eta>-1.5$).}
\label{tab:Bubble_fit_quality}
\end{table*}

\section{Additional results on the Falicov-Kimball model} \label{an:SR_FKM}

\begin{figure}[h]
    \centering
    \includegraphics[width=0.9\linewidth]{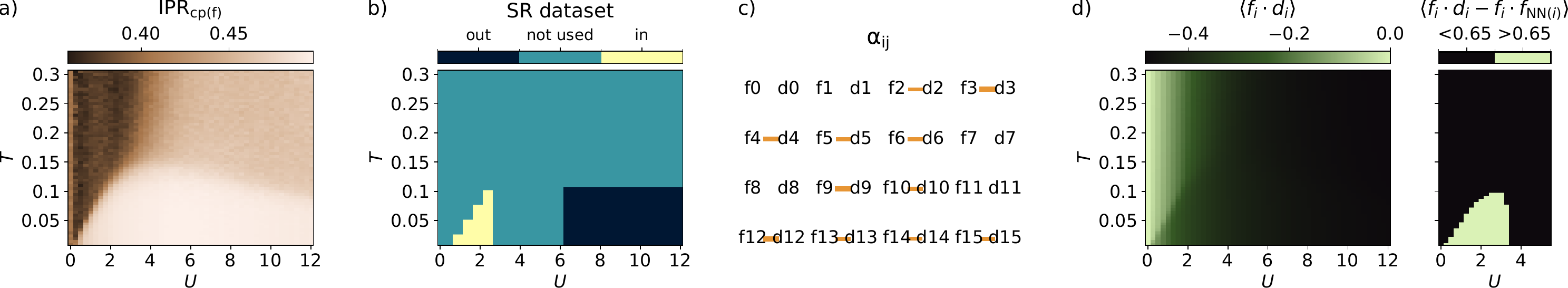}
    \caption{\textbf{Falicov-Kimball model: additional results.}
    \textbf{a)} Inverse participation ratio (IPR) computed from the conditional probabilities output by the decoder of the VAE. 
    \textbf{b)} Dataset used for the symbolic regression.
    \textbf{c)} Learned two-body correlator weights $\alpha_{ij}$. Links are colored according to the sign of $\alpha_{ij}$, orange being positive and purple negative, and their thickness scales with $|\alpha_{ij}|$
    \textbf{d)} Per-site particle correlator $\langle f_i\cdot d_i\rangle$ across the parameter space. Subtracting the nearest-neighbor correlator $f_i\cdot f_{\mathrm{NN}(i)}$ enables us to isolate the additional cluster (right).} 
    \label{fig:FKM_SR}
\end{figure}

In this appendix, we present complementary results on the Falicov-Kimball model to those presented in \cref{sec:fkm}. One advantage of having access to the conditional probabilities output by the decoder is the possibility of computing complex observables, sometimes inaccessible due to poor statistics on the gathered data. For instance, we use here the conditional probabilities to compute the inverse participation ratio (IPR),
\begin{equation}
\mathrm{IPR}_{cp(f)} = \frac{1}{N} \sum_i p(f_i \mid f_{j<i}, d_{j<i})^2,
\end{equation}
shown in~\cref{fig:FKM_SR}a. As a measure of localization, this quantity provides a clear signature of the weakly localized regime. Importantly, such an observable is hard to estimate from raw data alone. These results highlight that modeling conditional probabilities with the VAE not only captures nontrivial inter-species correlations that can be directly visualized, but also provides access to spatial localization properties that would otherwise remain hidden in the raw data.

Then, we present the symbolic regression results characterizing the variation of $|\mu_2|$ on the left side of the ordered phase. Already suggested by looking at the conditional probabilities, this behavior may result from $\mu_2$ capturing the repulsion strength of the $f$ and $d$ particles. To further investigate the nature of this additional cluster, we employ the SR classification scheme. In particular, we use a search space that includes all two-body interaction terms $\alpha_{ij}$ and add a learnable constant term to allow an overall shift in prediction. The parameter regions where data are labeled as inside/outside are illustrated in~\cref{fig:FKM_SR}b. The learned coefficients $\alpha_{ij}$ of the resulting two-body correlator are shown in~\cref{fig:FKM_SR}c. For clarity, only couplings with $|\alpha_{ij}|>0.35$ are displayed.

Remarkably, the learned structure contains exclusively on-site couplings between $f$ and $d$ particles. This observation naturally suggests considering the local correlator $\langle f_i \cdot d_i \rangle$. For repulsive interactions $U>0$, the expected correlator $\langle f_i \cdot d_i \rangle$ is always negative, reflecting the energetic penalty associated with simultaneous occupation of a site by both particle species. Varying $U$ effectively tunes the strength of this penalty, and correspondingly we observe that $\langle f_i \cdot d_i \rangle$ decreases monotonically with increasing $U$ (see left panel of~\cref{fig:FKM_SR}d). When combined with the $f_i\cdot f_{\mathrm{NN}(i)}$ correlator characterizing the ordered phase, it isolates the left side of the ordered phase, as illustrated in the right panel of~\cref{fig:FKM_SR}d. As concluded in the main text, the additional cluster therefore does not correspond to a distinct thermodynamic phase, but rather subdivides the ordered phase according to the degree to which joint $f$--$d$ occupancy is suppressed.

\end{document}